\documentclass[12pt, draftclsnofoot, onecolumn]{IEEEtran}	% regular single-column

 \usepackage{amsmath,amssymb}
 \usepackage{subfigure}
 \usepackage{graphicx,graphics,color,psfrag}
 \usepackage{cite,balance}
 \usepackage{algorithm}
 \usepackage{accents}
 \usepackage{amsthm}
 \usepackage{bm}
 \usepackage{url}
 \usepackage{algorithmic}
 \usepackage[english]{babel}
 \usepackage{multirow}
 \usepackage{enumerate}
 \usepackage{cases}
 \usepackage{stfloats}
 \usepackage{dsfont}
 \usepackage{color,soul}
 \usepackage{amsfonts}
 \usepackage{cite,graphicx,amsmath,amssymb}
 \usepackage{subfigure}
 \usepackage{fancyhdr}
 \usepackage{hhline}
 \usepackage{graphicx,graphics}
 \usepackage{array,color}
 \usepackage{amsmath}
 \usepackage{amsthm}

\newtheorem{theorem}{Theorem}

\newtheorem{definition}{Definition}

\newtheorem{lemma}{Lemma}
\newtheorem{corollary}{Corollary}

\newtheorem{assumption}{Assumption}

\newtheorem{remark}{\bf Remark}
\def\phi{\varphi}

\def\l{\left}
\def\r{\right}
\def\({\left(}
\def\){\right)}

\def\bb{{\mathbf{b}}}

\def\bz{{\mathbf{z}}}
\def\b0{{\mathbf{0}}}

\def\bA{{\mathbf{A}}}
\def\bB{{\mathbf{B}}}

\def\bF{{\mathbf{F}}}

\def\bH{{\mathbf{H}}}
\def\bI{{\mathbf{I}}}

\def\bN{{\mathbf{N}}}

\def\bQ{{\mathbf{Q}}}

\def\bS{{\mathbf{S}}}
\def\bT{{\mathbf{T}}}
\def\bU{{\mathbf{U}}}
\def\bV{{\mathbf{V}}}
\def\bW{{\mathbf{W}}}
\def\bX{{\mathbf{X}}}
\def\bY{{\mathbf{Y}}}
\def\bZ{{\mathbf{Z}}}

\newcommand{\tr}{\mathrm{tr}}

\newcommand{\nn}{\nonumber}

\begin{document}

\title{\huge Automatic Recognition of Space-Time Constellations  by Learning on the Grassmann  Manifold}
\author{Yuqing Du, Guangxu Zhu, Jiayao Zhang,  and Kaibin Huang
\thanks{ Y. Du, G. Zhu, and K.~Huang are with the Dept. of Electrical and Electronic Engineering and J. Zhang the Dept. of Computer Science, both at The University of Hong Kong, Hong Kong. Corresponding author: K. Huang (Email: huangkb@eee.hku.hk).}

}
\maketitle

\vspace{-15mm}
\begin{abstract}
Recent breakthroughs  in machine learning especially artificial intelligence  shift the paradigm of wireless communication towards \emph{intelligence radios}.  One of their  core operations is \emph{automatic modulation recognition} (AMR). Existing  research  focuses on coherent modulation schemes  such as  QAM, PSK and FSK. The  AMR of (non-coherent) space-time modulation remains an uncharted area despite its wide deployment in modern \emph{multiple-input-multiple-output} (MIMO) systems.  The scheme using a so called Grassmann constellation (comprising unitary matrices) enables rate-enhancement using multi-antennas and blind detection. In this work, we propose an AMR approach for Grassmann constellation based on data clustering, which differs from traditional  AMR  based on classification using a modulation database. The approach allows algorithms for clustering on the Grassmann manifold (or the Grassmannian), such as \emph{Grassmann K-means} and \emph{depth-first search} (DFS), originally developed for computer vision to be applied to AMR. We further develop an analytical framework for studying and designing these algorithms in the context of AMR. First, the \emph{maximum-likelihood} (ML) Grassmann constellation detection is proved to be equivalent to clustering on the Grassmannian. Thereby, a  well-known machine-learning  result that  was originally established   only for the Euclidean space is rediscovered for the Grassmannian.  Next, despite a rich literature on algorithmic design, theoretical analysis of data clustering is largely overlooked due to the lack of tractable techniques. We tackle the challenge by introducing   probabilistic metrics for measuring  the inter-cluster separability and intra-cluster connectivity of received space-time symbols and deriving them using tools from differential geometry and Grassmannian packing. The results provide useful insights into the effects of various parameters ranging from the signal-to-noise ratio to constellation size, facilitating algorithmic design. 

\end{abstract}

\vspace{-10pt}\section{introduction}
Recent breakthroughs in machine learning has motivated researchers to apply the technology to  the design of \emph{intelligent radios} for automating communication systems so as to simplify  their architectures or improve their  performance.  For instance, statistical learning has been used to merge channel estimation and data detection \cite{prasad, wen,  zhu}.  Moreover, it is also believed that radios with artificial intelligence can solve the  long-standing challenge of spectrum scarcity~\cite{Darpa}. Recent research trends in intelligent radios led to the revival  of the classic areas of \emph{cognitive radios} and  \emph{software defined radios} (SDR) \cite{haykin2005cognitive} focusing on leveraging machine learning to  attain a higher level of intelligence. In the areas of SDR or intelligent receivers, one important problem is \emph{automatic modulation recognition} (AMR), where a receiver blindly detects the modulation type and order of the received signals. This problem is challenging due to many unknown parameters at the receiver such as the signal power, carrier frequency-and-phase offsets, and timing as well as channel hostility. In the last two decades, extensive research has been conducted on AMR for  \emph{linear} and \emph{coherent}  modulation schemes (such as BPSK, QPSK, and QAM) and frequency-shift keying~\cite{azzouz2013automatic,dobre2007survey}. Interestingly, there exists little AMR technique for \emph{nonlinear} and \emph{non-coherent}  space-time modulation (or called Grassmann modulation) despite the extensive deployment in  \emph{multiple-input-multiple-output} (MIMO) systems. Grassmann modulation has emerged to be a promising solution for low-latency machine-type communication as it enables blind detection without \emph{channel state information} (CSI)   and high data rates~\cite{hochwald,durisi2016short}. This motivates the current work on filling the void of the area by  developing a novel  AMR approach for  Grassmann modulation, which will find applications in next-generation multi-antenna intelligent radios.

\vspace{-10pt}\subsection{Related Work and Motivation}

\subsubsection{Grassmann Modulation} Developed for MIMO systems, the modulation scheme  features a constellation consisting a set of subspace matrices embedded in the space-time signal space. Mathematically, the matrices are points on a Grassmann manifold, giving the name \emph{Grassmann constellation}.   The idea of Grassmann  modulation  was originally proposed in~\cite{hochwald,hochwald2} for achieving a linear growth of data rate with respect to the array sizes and the feature of blind symbol detection without CSI.  The feature results from the invariance of a Grassmann  modulated symbol (an orthonormal matrix) to MIMO channel rotation, which gives the technology an alternative name of   \emph{non-coherent MIMO}. Extensive research in this area focuses on designing practical Grassmann constellations including Fourier based~\cite{hochwald2} and  hierarchical designs~\cite{gohary2009} for efficient constellation generation, differential modulation for coping with fast fading~\cite{hochwald, hughes2000differential}, and  error probability minimization~\cite{mccloud2002signal}. From the information-theoretic perspective, the capacity of a MIMO channel with Grassmann modulation  was studied in~\cite{zheng2002communication}. A key finding is that  the capacity maximizing    constellation  is a solution of  subspace packing on the Grassmannian. 

Recent years have seen the resurgence of research interests on developing Grassmann modulation for next-generation wireless systems. The main reason is that its CSI-free feature makes it a promising solution for tackling the key challenges of reducing CSI overhead~\cite{love2008overview} and latency as   faced by many next-generation technologies including massive MIMO using large-scale arrays \cite{yang2013capacity}, full-duplex relaying \cite{gohary2014grassmannian}, and ultra-fast short-packet machine type communications \cite{durisi2016short}.  In view of its applications in future systems, it is thus important to consider Grassmann modulation in intelligent receiver design.

\subsubsection{Automatic Modulation Recognition}

The principle design approach adopted in existing AMR algorithms is \emph{classification} that maps the received signal to an element of a modulation database  combining different modulation types and orders~\cite{azzouz2013automatic}. The algorithms can be separated into two groups based on two typical mapping criteria, namely \emph{likelihood function} and \emph{feature distance} \cite{dobre2007survey}.  In the presence of \emph{additive white Gaussian noise} (AWGN) and  given a set of signal samples, a likelihood based algorithm typically computes  a likelihood function for each  modulation scheme in the database and then selects the most likely scheme used for modulating the signal (see e.g., \cite{wei2000maximum,beidas1995higher}). Though operating in a similar way,  a feature-based algorithm instead  computes the  feature vector of a modulated signal based on its  distribution  cumulants and then measures its vector distance   to   each modulation scheme (see e.g., \cite{swami2000hierarchical}).  

For feature-based AMR, the signal features derived from  cumulants  are design choices and may not be optimal especially for channels more complex than the   AWGN channels.  This motivates researchers to apply machine learning to train the modulation classifiers for improving the AMR accuracy~\cite{aslam2012automatic,west2017deep,hassan2012blind}. Specifically, in \cite{aslam2012automatic}, a hierarchical AMR algorithm was proposed that integrates  \emph{genetic programming}  (GP) and   the \emph{K-nearest-neighbour} algorithm, both of which are classic machine learning techniques. Furthermore, a  \emph{deep neural network}   was applied in  \cite{west2017deep} to AMR. For transmission over a MIMO channel, the received signal mixes a number of spatially multiplexed symbols, which increases the difficulty of AMR. It has been proposed in \cite{hassan2012blind} that the challenge can be tackled using an  \emph{artificial intelligent network}. 

Interestingly, though  Grassmann  modulation has been extensively studied and implemented in MIMO systems as discussed in the sequel, there exists no relevant AMR technique targeting the scheme. One possible reason is that existing designs cannot be straightforwardly extended to the Grassmann modulation due to its unique manifold structure. To be specific, existing AMR algorithms differentiate modulation schemes essentially by exploiting the statistical properties of a signal waveform in terms of  phase, magnitude and frequency. This approach is suitable for signal reception using a single antenna but is insufficient  for  MIMO transmission. For a MIMO receiver, matrix based properties of array observations  arise and it is important to exploit such properties  in  AMR. In particular, Grassmann modulated symbols are \emph{orthonormal matrices} that are mathematically  points on a Grassmannian embedded in the space-time  signal space.  How to exploit the unique manifold structure of Grassmann modulation in AMR remains an unexplored but important issue for its  relevance to  next-generation intelligent MIMO receivers.  

From the perspective of intelligent radios, the classic AMR algorithms lack the desired intelligence and flexibility. To be specific,  most algorithms involve a search over a modulation database comprising a set of combinations of modulation types and orders \cite{azzouz2013automatic,dobre2007survey}. It is impractical to include all possible combinations in the database as the required computing complexity is overwhelming.  As the result, the recognition capability of a receiver is limited by the modulation database, which is a drawback of the classic AMR approach. The rapid advancement in unsupervised learning calls for the development of  a modern intelligent AMR approach without the need of pre-specifying modulation types and orders. 

\vspace{-10pt}\subsection{Contributions}
In this work, we attempt to fill a  void in the AMR area by investigating  automatic  recognition of Grassmann modulation, referred to as  \emph{Grassmann AMR}. Specifically, the current work establishes a novel approach of Grassmann AMR based on data clustering on the Grassmannian via bridging the two areas of Grassmann AMR and unsupervised learning. Grassmann clustering algorithms  were originally developed for computer vision (see e.g., \cite{turaga}) and this is the first attempt on applying them to Grassmann AMR to the best of authors' knowledge. In the presence of channel noise, received Grassmann modulated symbols  form clusters on the Grassmannian with corresponding codewords as their centers.  Thus, it is a natural approach to apply manifold clustering techniques for AMR. Nevertheless, understanding its optimality and performance is challenging but important for guiding algorithmic design. This motivates the current work whose main contributions are summarized as follows. 

The first contribution of this work is to identify  the connection between \emph{maximum-likelihood (ML) detection} of Grassmann modulation  and data clustering on the Grassmannian.  To this end, we formulate the problem of ML constellation  detection and consider the well-known  \emph{expectation-maximization} (EM) algorithm for solving the problem.  The algorithm iterates between two steps, called the \emph{E-step} and the \emph{M-step}, till it converges. Under the assumption on high \emph{signal-to-noise ratio} (SNR), it is proved that the E-step is equivalent to projecting a   block of received symbols onto  the Grassmann manifold and clustering the projections using a given initial or updated Grassmann constellation. On the other hand, it is further proved that the M-step is equivalent to inferring the Grassmann  constellation via computing the centroids of the clusters of projected symbols. Combining the two equivalent steps is in fact the well-known  \emph{Grassmann K-means} algorithm in computer vision \cite{turaga}. The connection establishes the optimality of the proposed low-complexity AMR approach. From the perspective of learning, the result represents a significant finding  that the well-known connection between ML detection and data clustering originally known only for the linear Euclidean space  \cite{bishop2006pattern} also holds  on the non-linear  Grassmannian.

The second contribution is to analyze the performance of the proposed approach of Grassmann constellation detection  by data clustering. The developed framework not only yields  theoretic insights useful for designing  Grassmann AMR,  but also fills the void of the data-clustering area that lacks tractable performance analysis   \cite{turaga, bishop2006pattern}.  Specifically, we consider the K-means and \emph{depth-first search} (DFS) algorithms for constellation detection with and without prior knowledge of constellation size, respectively. The performance of both algorithms depends on the separability of clusters in the  dataset (the set of received symbols) and furthermore that of DFS requires the intra-cluster connectivity. To measure these dataset characteristics, suitable probabilistic  metrics are defined and  analyzed  by developing novel techniques such as ``Grassmannian bin packing" (see Fig.~\ref{fig_cluster_bin}) for analyzing intra-cluster connectivity.  These techniques leverage results  from differential geometry and subspace packing~\cite{dai2008quantization}. The derived results  quantify   the effects of various parameters on the detection performance, such as  the SNR, constellation and  dataset sizes, space-time dimension, and the DFS threshold. 

The last contribution of the work addresses the issue of how to embed a symbol-and-bit mapping in a Grassmann constellation so as to enable a receiver to detect bits following the blind symbol-and-constellation detection. A simple method is proposed that assigns ordered bit sequences to constellation codewords following the order of their subspace distances to a reference matrix, which is designed to be a truncated Fourier matrix.

%The remainder of the paper is organized as follows. Section~\ref{Grassmannian maths} provides some basic concepts on  Grassmann manifold. Sections~\ref{model} and~\ref{problem formulation} introduce the system model and problem formulation,  respectively. The equivalence  between Grassmannian data clustering and the EM algorithm for joint symbol-and-constellation detection is proved in Section~\ref{bridge}. Two clustering algorithms, K-means and DFS, are discussed  in Section~\ref{Algorithm_constellation_detection}. Section~\ref{analysis: DFS} provides the performance analysis of Grassmannian AMR by data clustering. Section~\ref{bit-symbol-error mapping} introduces a novel \emph{bit-symbol} mapping scheme allowing embedding in a Grassmann constellation. Simulation results are presented  in Section \ref{simulation}, followed by concluding remarks in Section~\ref{Section:Conclusion}.

\vspace{-10pt}\section{Mathematical Preliminaries}\label{Grassmannian maths}
To facilitate the subsequent exposition,  several basic  concepts and definitions related to Grassmann manifolds are introduced in this section.
\vspace{-10pt}\subsection{Stiefel and Grassmann Manifolds}
The  $(n, m)$ Stiefel manifold is the set of all $n$-by-$m$ orthonormal matrices for $1\leq m \leq n$, denoted by ${\cal{T}}_{n,m}$. Mathematically, the Stiefel manifold can be defined  as follows:
\begin{equation}
{\cal{T}}_{n,m} = \{ {\boldsymbol\Psi}\in\mathbb{C}^{n\times{m}}:  {\boldsymbol\Psi}^{H} {\boldsymbol\Psi} = \bI_{m} \}\label{eq:def 1}.
\end{equation}
On the other hand, the $(n, m)$  Grassmann manifold is a set of all $m$-dimensional subspaces in $\mathbb{C}^{n}$, denoted by ${\cal{G}}_{n,m}$. The manifold can be seen as the \emph{quotient space} of ${\cal{T}}_{n,m}$. To be specific, a point on the Grassmann manifold corresponds to  a class  of $n$-by-$m$ orthonormal matrices on the Stiefel manifold that span the same column   subspace defined by the point. Choose an arbitrary matrix $\boldsymbol\Upsilon$ from this class, called a \emph{generator}. Then the class, denoted as $[\boldsymbol\Upsilon]$, can be mathematically written as 
\begin{align}\label{representation}
[\boldsymbol\Upsilon] = \{\boldsymbol\Upsilon \bQ: \bQ\in {\cal O}_{m} \}.
\end{align}
where ${\cal O}_{m}$ denotes  the group of  $m\times m$ unitary matrices. The said  relation    between the Grassmannian ${\cal{G}}_{n,m}$ and the Stiefel  ${\cal{T}}_{n,m}$ is typically represented by ${\cal{G}}_{n,m} = {\cal{T}}_{n,m}/{\cal O}_{m}$. Based on this relation and the definition of the class $[\boldsymbol\Upsilon]$ in \eqref{representation}, a Grassmann point mapped to this class can be then represented by the generator $\boldsymbol\Upsilon$ for ease of notation. 

\begin{figure} [tt]
\centering
\includegraphics[width = 9cm]{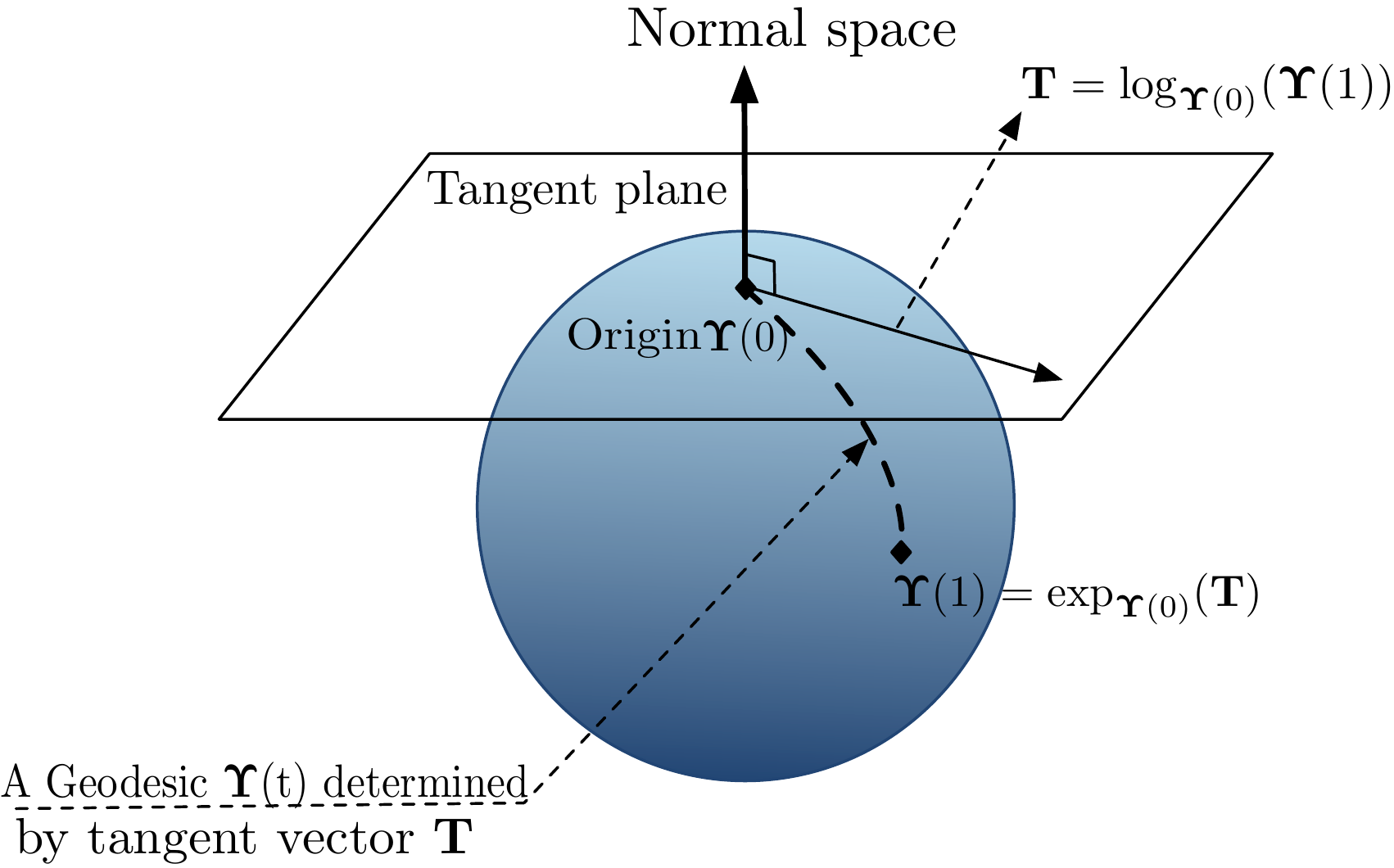}
\caption{A Grassmann manifold and related subspaces and mappings. }
\label{Grassmann manifold}
\end{figure}

\vspace{-10pt}\subsection{Tangent and Normal Spaces of Grassmann Manifold}

To perform differential calculus on a manifold, one needs to specify its tangent and normal spaces. As illustrated in Fig. \ref{Grassmann manifold}, for each point $\boldsymbol\Upsilon$ on the Grassmann manifold, there exists a \emph{tangent space}, referred to  the hyperplane tangent to the manifold at $\boldsymbol\Upsilon$ and having the same dimensions as that of the  manifold. For any vector $\boldsymbol\Delta$ in the tangent space, it satisfies $\boldsymbol\Upsilon^{H}\boldsymbol\Delta = \b0$. A \emph{normal space} with respect to a given tangent space is defined to be the orthogonal complement of the latter. For  each vector  ${\bN}$ in a  normal space, it can be represented as $\bN =  {\boldsymbol\Upsilon}\bS$, where ${\boldsymbol\Upsilon}$ is the point of tangency on the Grassmann manifold and  $\bS$ is some $m$-by-$m$ symmetric matrix.

\vspace{-10pt}\subsection{Geodesics on Grassmann Manifold}
Roughly  speaking, a geodesic is the shortest curve linking two points on a Grassmannian   as illustrated in Fig. \ref{Grassmann manifold}. By representing  the geodesic as a function  $\boldsymbol\Upsilon(t)$ with $|t|\leq 1$, its two end points are $\boldsymbol\Upsilon(0)$ and $\boldsymbol\Upsilon(1)$. An important property of  geodesics on a Grassmannian  is given as follows.
\begin{lemma}[\cite{edelman}]\label{Geodesic Property: lemma 1}
\emph{For any geodesic $\boldsymbol\Upsilon(t)$ on a Grassmannian, it must satisfy the following equation:
\begin{equation}
\ddot{{\boldsymbol\Upsilon}} + {\boldsymbol\Upsilon(t)}({\dot{{\boldsymbol\Upsilon}}}^{H}\dot{{\boldsymbol\Upsilon}}) = 0,
\end{equation}
where $\dot{{\boldsymbol\Upsilon}} = {d{\boldsymbol\Upsilon(t)}}/{dt}$ is the \emph{velocity vector} and $\ddot{{\boldsymbol\Upsilon}} = {d^{2}{\boldsymbol\Upsilon}}(t)/{dt^{2}}$ is the \emph{acceleration vector}. The  vectors $\dot{{\boldsymbol\Upsilon}}$ and $\ddot{{\boldsymbol\Upsilon}}$ lie in the tangent and normal space of the manifold, respectively.}
\end{lemma}

\vspace{-5mm}
\vspace{-10pt}\subsection{Exponential and Logarithm Mappings}

\begin{definition}[Exponential Mapping~\cite{edelman}]\label{Gram: exp mapping}
\emph{As illustrated  in Fig.~\ref{Grassmann manifold}, The exponential mapping, denoted by $\exp_{\boldsymbol\Upsilon(0)}(t\bT) = \boldsymbol\Upsilon(t) $ with $|{t}|\leq{1}$, is a one-to-one mapping from a velocity vector $ t \bT = t\dot{\boldsymbol\Upsilon}(0)$ in  the tangent plane with the tangency at the point $\boldsymbol\Upsilon(0)$ to a point $\boldsymbol\Upsilon(t)$ on the Grassmannian. Mathematically, by denoting $\boldsymbol\Upsilon(0)$ as $\boldsymbol\Upsilon_{0}$ and decomposing $\bT$ by \emph{singular-value decomposition} (SVD) as  $\bT=\bU{\boldsymbol\Sigma}\bV^{H}$, the exponential mapping can be computed as
\begin{equation}
{\exp}_{\boldsymbol\Upsilon_{0}}(\bT) = (\boldsymbol\Upsilon_{0}{\bV}  \quad \bU)\begin{pmatrix}	\cos\boldsymbol\Sigma\\ \sin\boldsymbol\Sigma \end{pmatrix}\bV^{H}\label{eq:exp}. 
\end{equation}}
\end{definition}

\begin{definition}[Logarithm Mapping~\cite{edelman}]\label{lemma: logarithm mapping}
\emph{The logarithm mapping, denoted as $\log_{\boldsymbol\Upsilon(0)}{\boldsymbol\Upsilon(t)} = t\bT$ with $|{t}|\leq{1}$, is  the \emph{inverse exponential mapping} and maps a point  on the Grassmann manifold back to the corresponding velocity vector. Mathematically, given  two points $\bA$ and $\bB$  on the Grassmann manifold, the logarithm mapping that generates a velocity vector $\bT$ pointing from  $\bA$ to $\bB$ can be computed as 
\begin{equation}\label{eq:log_orig}
\log_{\boldsymbol\bA}{\bB} = \bT = \bU{\boldsymbol\Sigma}\bV^{H},
\end{equation}
where the SVD  components $\bU$, $\bV$ and $\boldsymbol\Sigma$ can be obtained via the  \emph{cosine-sine decomposition}: 
\begin{equation}
 \begin{pmatrix} \bV (\cos{\boldsymbol\Sigma})\bV^{H}\\\bU (\sin{\boldsymbol\Sigma}) \bV^{H}\end{pmatrix} = \begin{pmatrix}	\bA^{H}\bB\\(\bI-\bA{\bA}^{H})\bB \end{pmatrix}. \label{eq:log}
\end{equation}
}
\end{definition}

\vspace{-10pt}\section{System Model}\label{model}
Consider a point-to-point MIMO system comprising a pair of multi-antenna transmitter and receiver. The numbers of transmit and receive antennas are denoted as  ${N}_{t}$ and ${N}_{r}$, respectively. It is assumed that $N_r$ is larger than $N_t$ so that the receiver can observe the space-time symbols. Time is slotted. Each space-time symbol occupies $T$ slots. The block-fading channel model is adopted, where the channel coefficients remain unchanged within a symbol duration and change independently over different durations. The $N_t\times N_r$ MIMO channel matrix $\bH$ comprises \emph{independent and identically distributed} (i.i.d.) $\mathcal{CN}(0,1) $ coefficients. Consider the $i$-th symbol duration in a block of $N$ space-time symbols. Let $\bX^{(i)}$ denote the transmitted space-time symbol that is a $T \times N_t$ matrix, $\bH^{(i)}$ the  channel matrix, and $\bY^{(i)}$ the $T \times N_r$ received symbol. For ease of notation, following \cite{hochwald2,gohary2009},  the baseband input-output relationship of the system can be written as 
\begin{equation}\label{eq: single_user_model}
\bY^{(i)}=\bX^{(i)}\bH^{(i)}+\sqrt{\frac{N_t}{\rho T}}\bW^{(i)}, \; i = 1,2,\cdots,N,
\end{equation}
where  $\rho$ represents the transmit SNR and $\bW^{(i)}\in\mathbb{C}^{T\times{N_r}}$  the AWGN comprising  i.i.d. $\mathcal{CN}(0,1)$ elements. 
\begin{assumption}[Receiver Knowledge]\label{asmp1}
\emph{The receiver has no knowledge of the Grassmann constellation used by the transmitter. However, the receiver has information on the size of the transmit array, $N_t$, the symbol  duration $T$ and symbol boundaries so as to receive the symbol block  $\{\bY^{(i)}\}$ in \eqref{eq: single_user_model}.\footnote{Under the assumption of $N_r \geq N_t$, $N_t$ can be estimated  by observing the ranks of received data symbols. For receiver synchronization, the symbol duration and boundaries can be estimated using standard methods in the literature (see e.g.,  \cite{azzouz2013automatic}).}}
\end{assumption}

Transmitted symbols $\{\bX^{(i)}\}$ are modulated using a Grassmann constellation codebook, denoted as $\mathcal{F}$. On the other hand,   the codebook detected by the receiver is denoted as $\hat{\mathcal{F}}$. To combat fading and enable non-coherent detection without CSI, the $T\times N_t$ modulated symbols are designed to be ``tall" matrices with $T\geq N_t$. Consequently, information is embedded in the column space of each symbol. It is important to note that given tall symbol matrices,  propagation over the MIMO channel changes only the symbol's row space but not its column space. Therefore,  the symbols $\{\bX^{(i)}\}$ can be detected at the receiver  by  computing the  column spaces of received symbols $\{\bY^{(i)}\}$ without requiring CSI~\cite{hochwald,hochwald2}. For consistency in matrix notation, let the Grassmann codebook $\mathcal{F}$ be a set of $T\times N_t$ tall orthonormal matrices, called \emph{codewords}: $\mathcal{F} = \{\boldsymbol\mu_{\ell}\}$ with $\boldsymbol\mu_{\ell} \in\mathbb{O}^{{T}\times{N_t}}$, where $\mathbb{O}$ represents the group of orthonormal matrices.  
%Then a modulated symbol is the transpose of the corresponding  codeword. 

From the perspective of communication performance, it is well known that it is desirable to maximize the  pairwise distances between elements of the constellation $\mathcal{F}$. In other words,  the optimal constellation design is related to the following problem of subspace packing \cite{conway,love}:
\begin{equation}\label{Eq:Packing}
\text{(Subspace Packing)} \quad  \max_{\mathcal{F}\subset\mathcal{G}} \min_{\ell\neq{n}} d(\boldsymbol\mu_{\ell},\boldsymbol\mu_{n}),  
\end{equation}
where $d(\cdot,\cdot)$ is a subspace  distance metric. Among many others, two commonly used metrics are considered in this paper, namely \emph{geodesic distance}, denoted as $d_g(\cdot, \cdot)$ and \emph{Procrustes distance}, denoted as $d_p(\cdot, \cdot)$. Given two points $\boldsymbol\Upsilon$ and $\boldsymbol\Upsilon'$  on the Grassmannian, $d_g(\boldsymbol\Upsilon, \boldsymbol\Upsilon')$ measures the length of the geodesic and $d_p(\boldsymbol\Upsilon, \boldsymbol\Upsilon')$ the Euclidean distance between them:  
\begin{align}
d_g(\boldsymbol\Upsilon, \boldsymbol\Upsilon') &= {\parallel{\log_{ \boldsymbol\Upsilon}(\boldsymbol\Upsilon')}\parallel}_F\label{eq:distance},\\
d^2_p(\boldsymbol\Upsilon, \boldsymbol\Upsilon') &= N_t - \text{tr}\l\{ \boldsymbol\Upsilon\boldsymbol\Upsilon^{H}\boldsymbol\Upsilon'(\boldsymbol\Upsilon')^{H}\r\}, \label{eq: procrustes_dist}
\end{align}
where  $\log_{ \boldsymbol\Upsilon}(\boldsymbol\Upsilon')$ is the logarithm mapping defined in \eqref{eq:log_orig} and $N_t$ denotes the dimension of the Grassmannian. Finding the optimal constellation by subspace packing is in general intractable and typically relies on numerical computation \cite{conway}. However, the computed constellation is not  unique, which further motivates the  assumption of unknown constellation at the receiver and the need of AMR.

\vspace{-10pt}\section{Problem Formulation}\label{problem formulation}
In this section, we first formulate the problem of ML symbol detection and then build on it to formulate  the problem of ML Grassmann  constellation detection. 

\vspace{-10pt}\subsection{Maximum-Likelihood  Symbol Detection}

Consider the  communication model  in \eqref{eq: single_user_model} and the assumed Gaussian distributions of channel and noise. Given  the transmitted symbols $\{\bX^{(i)}\}$ and no CSI, the received symbols  $\{\bY^{(i)}\}$ are i.i.d.  complex Gaussian random matrices whose conditional distribution is $\bY^{(i)}| \bX^{(i)} \thicksim {\cal{CN}}\(\b0,\bX^{(i)}(\bX^{(i)})^H +\frac{N_t}{\rho T}\bI_{T}\)$. Specifically, the  distribution is  given by  \cite{gohary2009}
\begin{align}
p(\bY^{(i)}|\bX^{(i)}) =  \frac{ \exp\(  -\frac{\rho T}{N_t} \text{tr} \(  (\bY^{(i)})^{H}( \bI_T - \frac{1}{1+N_t/\rho T}\bX^{(i)}(\bX^{(i)})^{H}         )\bY^{(i)}    \)     \) }{(\pi N_t/\rho T)^{TN_t}(1 + \rho T/N_t)^{N_tN_r}}\label{distribution} .
\end{align}
For the conventional case where the constellation codebook ${\mathcal{F}}^{*}$ is known  at receiver, the problem of ML symbol  detection 
can be mathematically formulated as (see e.g., \cite{gohary2009})
\begin{equation}\label{conventional ML}
 {\hat \bX}^{(i)} = \max_{\bX^{(i)}\in{\mathcal{F}}^{*}} p(\bY^{(i)}| \bX^{(i)}), \;\forall i.
\end{equation}
Based on  \eqref{distribution}, an equivalent problem is 
\begin{equation}\label{eq:traditional symbol detection}
{\hat \bX}^{(i)} = \arg \max_{{\bX}^{(i)} \in {\mathcal{F}}^{*} } \text{tr} \l\{ (\bY^{(i)})^{H}{\bX^{(i)}}(\bX^{(i)})^{H}\bY^{(i)} \r\}, \;\forall i.
\end{equation}

\vspace{-10pt}\subsection{Maximum-Likelihood  Constellation Detection}

For the current  case that the ground-true constellation ${\mathcal{F}}^{*}$ is unknown \emph{a priori}, we need to first infer ${\mathcal{F}}^{*}$ from the block of received symbols  $\bY = \{\bY^{(i)}\}_{i=1}^N$. To simplify exposition, even though ${\mathcal{F}}^{*}$ is unknown, its size, denoted as $L$, is assumed to be  known at the receiver. The issue of unknown constellation size at the receiver is addressed  in Sections~\ref{ms}. Then the   ML problem  formulation is 
\begin{equation}
\hat{\mathcal{F}} =\arg\max_{\mathcal{F}} \log p(\bY|{\mathcal{F}}) =\arg\max_{\mathcal{F}} \sum_{i=1}^{N}\log p(\bY^{(i)}|{\mathcal{F}}).\label{original detection problem:a}
\end{equation} 
The likelihood function $p(\bY^{(i)}|{\mathcal{F}})$ follows  the \emph{mixture of Gaussian} (MoG) model given by
\begin{equation}
p(\bY^{(i)}|{\mathcal{F}}) \!=\! \sum_{\ell} p(\bY^{(i)}|\bX^{(i)} = \boldsymbol\mu_{\ell}, {\mathcal{F}})p(\bX^{(i)} = \boldsymbol\mu_{\ell} | {\mathcal{F}}), \;\; \forall i\label{mixture}. \!\!
\end{equation}
To facilitate subsequent analysis, we introduce a new  latent variable $\bZ = [\bz_1, \dots, \bz_N]$ where $\bz_{i} = [z_{i,1}, z_{i,2}, \cdots, z_{i,L}]^T$ is a $L$-dimensional binary random vector indicating the index of codeword modulating  the $i$-th transmitted  symbol $\bX^{(i)}$. For instance, if $\{ \bX^{(i)} = \boldsymbol\mu_{\ell}\}$, we have $z_{i,\ell} = 1$ with  the remaining elements in  $\bz_{i}$ being  zeros. Due to the equivalence between the two events $\{ z_{i,\ell} = 1\}$ and $\{ \bX^{(i)} = \boldsymbol\mu_{\ell}\}$, the MoG model in \eqref{mixture} can be rewritten as
\begin{equation}
p(\bY^{(i)}|{\mathcal{F}}) = \sum_{\ell} p(\bY^{(i)}|\bz_{i,\ell} = 1, {\mathcal{F}})p(\bz_{i,\ell} = 1 | {\mathcal{F}}), \;\; \forall i \label{mixture_Z}.
\end{equation}
By substituting  \eqref{mixture_Z} into \eqref{original detection problem:a}, the problem of constellation detection is rewritten as 
\begin{equation}
\hat{\mathcal{F}} = \arg\max_{\mathcal{F}} \sum_{i=1}^{N}\log \sum_{\ell} p(\bY^{(i)}|\bz_{i,\ell} = 1, {\mathcal{F}})p(\bz_{i,\ell} = 1 | {\mathcal{F}}).
\label{original detection problem}
\end{equation} 
Directly solving  this  optimization  problem  is intractable  due to the \emph{non-convexity} of the objective function  arising from  the existence of the latent \emph{random variable} (r.v.) $\bZ$ (or equivalently  the symbols $\{\bX^{(i)}\}$). A commonly used approach   for solving such a  non-convex ML problem with latent variables  is the EM algorithm as discussed in the following section.

\vspace{-10pt}\section{Grassmann Constellation Detection:  From EM to Data Clustering}\label{bridge}
In this section, we consider the application of the well-known EM algorithm for solving the problem of ML constellation detection formulated in the preceding section. The main task of this section is to prove the equivalence between the EM algorithm and the proposed detection approach of data clustering on the Grassmannian. 

\vspace{-10pt}\subsection{Grassmann Constellation Detection by EM }
\subsubsection{Implementation of EM}
Consider the problem of ML estimation of the codebook $\mathcal{F}$ based on  the observation $\bY$ and given a latent variable $\bZ$. 
The EM algorithm for solving  the  problem specified in \eqref{original detection problem} iterates between the two main steps \cite{bishop2006pattern}: 
\begin{align}
{(\bf E\!-\!step):} \;\;&\text{Evaluate} \quad  p(\bZ|\bY,{\hat {\mathcal{F}}}) = \prod_{i = 1}^{N}\prod_{\ell = 1}^{L} r_{i,\ell}^{z_{i,\ell}}\label{E-step},\\
{(\bf M\!-\!step):}\;\; &\text{Solve} \quad  {\hat {\mathcal{F}}} = \arg \max_{{\mathcal{F}}} \mathbb{E}_{\bZ}[\log p(\bY,\bZ|{\mathcal{F}})]\label{M-step},
\end{align}
where we define $r_{i,\ell} = p(z_{i,\ell} = 1|\bY^{(i)},{\hat {\mathcal{F}}})$. For the E-step in  \eqref{E-step}, the posterior distribution of the latent variable $\bZ$ is calculated using the current estimation of the  codebook   $\hat{\mathcal{F}}$, where the calculation involves evaluating the set of variables $\{r_{i,\ell}\}$. For the M-step in \eqref{M-step}, the codebook $\hat{\mathcal{F}}$ is updated by maximizing the expectation of the complete-data log-likelihood, which can be evaluated using 
 the posterior distribution updated in the E-step as follows: 
\begin{align}
\mathbb{E}_{\bZ}[\log p(\bY,\bZ|{\mathcal{F}})] & = \sum_{\bZ} p(\bZ|\bY,\mathcal{F})\log p(\bY,\bZ|{\mathcal{F}})\\
& =  \sum_{\bZ} p(\bZ|\bY,{\mathcal{F}})\log \l(p(\bY|\bZ,{\mathcal{F}})p(\bZ)\r)\label{M_step_ril}.
\end{align}
The  specific expressions of the E-step and M-step can be derived as follows. For ease of notation, denote $\pi_{\ell} = p(z_{i,\ell}=1)$. It follows that $p(\bZ) = \prod_{i=1}^{N}\prod_{\ell=1}^{L}\pi^{z_{i,\ell}}_{\ell}$ and
 $p(\bY|\bZ,{\mathcal{F}}) = \prod_{i=1}^{N}\prod_{\ell=1}^{L}p(\bY^{(i)}|\bX^{(i)}=\boldsymbol\mu_{\ell},{\mathcal{F}})^{z_{i,\ell}}$.
 Substituting them into \eqref{M_step_ril} and following the standard procedure in \cite[Section 9.3]{bishop2006pattern}, the E-step variables  $\{r_{i,\ell}\}$ and  $\mathbb{E}_{\bZ}[\log p(\bY,\bZ|{\mathcal{F}})]$ for the M-step   are given by:
\begin{align}\label{soft_assignment}
r_{i,\ell} = \frac{ \pi_{\ell} p(\bY^{(i)}|\bX^{(i)}=\hat{\boldsymbol\mu}_{\ell},\hat{\mathcal{F}}) }{ \sum_{j = 1}^{L} \pi_{j}p(\bY^{(i)}|\bX^{(i)}=\hat{\boldsymbol\mu}_{j},\hat{\mathcal{F}})},
\end{align} 
\begin{equation}\label{M-step v2}
 \mathbb{E}_{\bZ}[\log p(\bY,\bZ|{\mathcal{F}})] = \sum_{i=1}^{N}\sum_{\ell=1}^{L} r_{i,\ell} (\log \pi_{\ell} +  \log p(\bY^{(i)}|\bX^{(i)}=\boldsymbol\mu_{\ell},{\mathcal{F}})).
\end{equation}

Note that the probability $r_{i,\ell}$ can be interpreted as a \emph{soft assignment} of the $i$-th received symbol $\bY^{(i)}$ to the $\ell$-th codeword $\hat{\boldsymbol\mu}_\ell$. Moreover, given the estimated $\{r_{i,\ell}\}$ and using  \eqref{distribution}, one can show that maximizing \eqref{M-step v2} in the M-step is equivalent to  maximizing $\sum_{i=1}^{N}\sum_{\ell=1}^{L}r_{i,\ell} \text{tr}\l\{ ({\bY^{(i)}})^{H}\boldsymbol\mu_{\ell}\boldsymbol\mu^{H}_{\ell}\bY^{(i)}  \r\}$. Thereby, the EM algorithm for Grassmann constellation detection can be implemented  as:
\begin{align}
{(\bf E\!-\!step):}  \;&\text{Evaluate} \;\; \{r_{i,\ell}\} \ \text{using \eqref{soft_assignment}.} \label{E_step_temp}\\
{(\bf M\!-\!step):}\;&\text{Solve} \;\; {\hat {\mathcal{F}}} =\arg\max_{{\mathcal{F}}} \sum_{i=1}^{N}\sum_{\ell=1}^{L}r_{i,\ell} \text{tr}\l\{ ({\bY^{(i)}})^{H}\boldsymbol\mu_{\ell}\boldsymbol\mu^{H}_{\ell}\bY^{(i)}  \r\}\label{M_step_temp}.
\end{align}

\subsubsection{Difficulties of EM Implementation}\label{convergence speed}
The direct application of  the EM algorithm faces two main  difficulties described as follows. 
\begin{itemize}
\item The  optimization problem in the M-step  in \eqref{M_step_temp} is \emph{non-convex} and thus difficult to solve. Specifically, the non-convexity is due to the maximimization of a convex object function  under the   constraints that  the codewords (variables)  $\{\boldsymbol\mu_{\ell}\}$ are subspace matrices or equivalently points on the Grassmannian. 
\item The convergence  for implementing the EM algorithm based on  the MoG model in \eqref{mixture} is potentially slow as the model  involves Gaussian components with overlapping means (that are all zeros). As  proved in \cite{xu1996convergence}, the convergence rate of the EM algorithm on a MoG model is faster if the Gaussian components are better separated.
%\item  The  evaluation of the E-step could be computationally intensive due to the potential high-dimensional integral  in the posterior distribution calculation where the dimension is determined by the symbol block length that is usually large. 
\end{itemize}
To overcome these difficulties, we prove in the sequel the equivalence of  the EM algorithm with the Grassmann K-means algorithm, a widely used clustering algorithm. The latter algorithm has a faster convergence rate and lower complexity  due to the well-separated symbol  clusters  ``seen'' on the Grassmannian as revealed in Lemma~\ref{lemma: coverage of a typical cluster} in the sequel and the discussion therein.

\vspace{-10pt}\subsection{Asymptotic Equivalence between EM and Data Clustering}
In this sub-section, we prove that the EM algorithm for Grassmann constellation detection as derived in the preceding section is asymptotic equivalent to     data clustering on the Grassmannian when the transmit SNR is high and the dataset size $N$ is sufficiently large. The result allows  the replacement of the complex EM algorithm with the low-complexity clustering algorithms from machine learning. 

\subsubsection{From E-step to symbol detection} Consider the EM E-step in \eqref{E_step_temp}. First, substituting the conditional distribution of the received symbol  $\bY^{(i)}$ in \eqref{distribution} into the soft assignments $\{r_{i,\ell}\}$ in \eqref{soft_assignment} leads to the following result. 

\begin{lemma}\emph{(From Soft to Hard Assignments).}\label{Lem:HardAssign}
\emph{For a high  transmit SNR ($\rho \rightarrow \infty$), the soft assignments of received symbols, $\{r_{i,\ell}\}$,  become \emph{hard assignments}  taking only binary values:}
\begin{align}\label{eq: hard assignment}
r_{i,\ell} \rightarrow  \left\{ 
 \begin{array}{rcl}
 1, &  & \ell = \arg\max\limits_{j} \tr\l\{ (\bY^{(i)})^{H}\hat{\boldsymbol\mu}_{j}(\hat{\boldsymbol\mu}_{j})^{H}\bY^{(i)} \r\};\\
 0, &  & \emph{\text{otherwise}},
\end{array}
\right. 
\end{align}
\emph{where $\bY^{(i)}$ is the $i$-th received symbol and $\hat{\boldsymbol\mu}_{j}$ the $j$-th codeword in the estimated codebook $\hat{\mathcal{F}}$.} 
\end{lemma}

Next, we can show that the hard assignments of symbols to codewords in Lemma~\ref{Lem:HardAssign} are approximately based on the criterion of \emph{shortest subspace distance}. To this end, define the $i$-th received Grassmann symbol  $\boldsymbol\Upsilon^{(i)}$ as the dominant  $N_t$ dimensions of the left  eigen-space of the received symbol $\bY^{(i)}$, which is its only SVD component containing   information on the transmitted symbol. Specifically, consider the following SVD of $\bY^{(i)}$
\begin{align}\label{decomposition_Y}
\bY^{(i)} &= \l[\begin{matrix} \bU^{(i)}_Y & \bU^{(i)}_W \end{matrix}  \r]		
 \l[\begin{matrix} 
  \boldsymbol{\Sigma}^{(i)}_Y & \b0 \\ 
\b0 & \boldsymbol{\Sigma}^{(i)}_W
\end{matrix}\r]
\l[\begin{matrix} 
 (\bV^{(i)}_Y)^{H}\\ 
(\bV^{(i)}_W)^{H}
\end{matrix}\r],
\end{align}
%\begin{align}
%\bY^{(i)} &= \l[\begin{matrix} \bU^{(i)} & \widetilde{\bU}^{(i)} \end{matrix}  \r]		
% \boldsymbol{\Sigma}^{(i)}  
%\l[\begin{matrix} 
% \bV^{(i)}\\ 
%\widetilde{\bV}^{(i)}
%\end{matrix}\r]
%\end{align}
where  the diagonal elements of $\boldsymbol{\Sigma}^{(i)}_Y$ and $\boldsymbol{\Sigma}^{(i)}_W$  are the $q = \min(N_r,T)$ singular-values $\sigma_1, \sigma_2, \cdots, \sigma_{q}$ arranged in the descending order, and $\bU^{(i)}_Y$ and  $(\bV^{(i)}_Y)^{H}$ are the dominant $N_t$ dimensional left and right eigen-subspace, respectively. Then the  Grassmann symbol (a tall matrix) is $\boldsymbol\Upsilon^{(i)} = \bU^{(i)}_Y$. 
%(XXX: I changed the transpose notation; please revise the following notations)
%XXX: I have revised the lemma below. I believe largest eigenvalue is better bound. 
\begin{lemma}\label{single bound fast fading}
\emph{The  hard assignment criteria  in Lemma \ref{Lem:HardAssign} can be bounded as follows: }
\begin{align}
\l(\sigma^{(i)}_{N_t}\r)^2\l[N_t - d^2_p\l(\boldsymbol\Upsilon^{(i)}, \hat{\boldsymbol\mu}_{j}\r)\r]     &\leq \tr\l\{ (\bY^{(i)})^{H}\hat{\boldsymbol\mu}_{j}(\hat{\boldsymbol\mu}_{j})^{H}\bY^{(i)}\r\}\\
& \leq \l(\sigma^{(i)}_1\r)^2\l[N_t - d^2_p\l(\boldsymbol\Upsilon^{(i)}, \hat{\boldsymbol\mu}_{j}\r)\r] ,    
\end{align}
\emph{where $\sigma^{(i)}_k$ denotes  the $k$-th singular value of the received symbol $\bY^{(i)}$, and $d_p(\cdot, \cdot)$ is the Procrustes distance defined in \eqref{eq: procrustes_dist}.}
\end{lemma}
The proof is presented in   Appendix~\ref{proof:assignment criteria}. Approximating the hard assignment criteria in Lemma \ref{Lem:HardAssign} by either the lower or the upper bound in Lemma \ref{single bound fast fading} leads to the following hard-assignment based on the Procrustes distance: 
\begin{align}\label{eq:DistanceAssign}
r_{i,\ell} \rightarrow  \left\{ 
 \begin{array}{rcl}
 1, &  & \ell = \arg\min\limits_{j}d^2_p\l(\boldsymbol\Upsilon^{(i)}, \hat{\boldsymbol\mu}_{j}\r);\\
 0, &  & \text{otherwise}.
\end{array}
\right. 
\end{align}
It  follows that  the E-step of the EM algorithm in \eqref{E_step_temp} can be approximated by the computation of the assignment variables $\{r_{i,\ell}\}$ using \eqref{eq:DistanceAssign}. As a result, the E-step is equivalent to clustering the received symbols using the estimated codewords $\{\hat{\boldsymbol\mu}_{j}\}$ and the criteria of shortest Procrustes  distance. Note that in the high SNR regime, one can infer from the system equation in \eqref{eq: single_user_model} that  the singular values of $\bY^{(i)}$ are approximately equal to those of the channel matrix $\bH^{(i)}$. Thus, when the channel is well conditioned $\l(\sigma^{(i)}_{N_t}\approx \sigma^{(i)}_1\r)$, the approximation of the E-step by \eqref{eq:DistanceAssign} is accurate.

\subsubsection{From M-step to codeword optimization} Consider the EM M-step in \eqref{M_step_temp}. For a sufficiently high SNR and a sufficiently large dataset size, it is proved in the sequel that the M-step is equivalent to codeword optimization. Specifically, each estimated codeword in the constellation codebook is updated by computing the Grassmann \emph{centroid}, which has the minimum sum subspace distances to  the cluster of estimated  Grassmann symbols associated with the codeword.

Consider a particular cluster of received symbols detected as the $\ell$-th codeword in the E-step. Their indices can be grouped in the set ${\mathcal C}_{\ell} = \{i\mid r_{i,\ell} = 1\}$ with the assignments $\{r_{i,\ell}\}$ given in Lemma \ref{Lem:HardAssign}. The number of symbols in ${\mathcal C}_{\ell}$ is denoted as $N_\ell = |{\mathcal C}_{\ell}|$. 
Consider the M-step in \eqref{M_step_temp}. Using the definition of the index set ${\mathcal C}_{\ell}$, the M-step can be rewritten as 
\begin{equation}
{\hat {\mathcal{F}}} =\arg\max_{{\mathcal{F}}} \sum_{\ell=1}^{L}\sum_{i\in \mathcal{C}_\ell}\text{tr}\l\{ ({\bY^{(i)}})^{H}\boldsymbol\mu_{\ell}\boldsymbol\mu^{H}_{\ell}\bY^{(i)}  \r\}.
\end{equation}
This is equivalent to optimizing the codewords as follows: 
\begin{equation}\label{Eq:BookOptim}
\hat {\boldsymbol{\mu}}_\ell =\arg\max_{\boldsymbol{\mu}_{\ell}\in \mathcal{G}} \sum_{i\in \mathcal{C}_\ell} \text{tr}\l\{ ({\bY^{(i)}})^{H}\boldsymbol\mu_{\ell}\boldsymbol\mu^{H}_{\ell}\bY^{(i)}  \r\}, \quad \forall \ell.
\end{equation}

Next, an asymptotic form of the above codeword  optimization is obtained for the case of large dataset size. To this end, define the minimum (pairwise) distance of the constellation codebook $\mathcal{F}$ as  
\begin{align}\label{min_distance}
d_{\min} = \min \limits_{\substack{\boldsymbol{\mu}, \boldsymbol{\mu}' \in \mathcal{F}\\ \boldsymbol{\mu}\neq \boldsymbol{\mu}'}}d_p(\boldsymbol{\mu}, \boldsymbol{\mu}').
\end{align}
\begin{lemma}\label{Lem: SufficientDataSize}
\emph{If the minimum distance of the codebook $\mathcal{F}$ is strictly positive and all codewords are transmitted with equal probabilities, as the symbol dataset size $N\rightarrow \infty$, the symbol cluster size $N_{\ell}\rightarrow \infty$ for all $\ell$. 
}
\end{lemma}
The proof is presented in   Appendix \ref{proof: SufficientDataSize}.  Using the result and applying the law of large numbers, we can obtain the following important asymptotic form of the summation term in \eqref{Eq:BookOptim}. 
\begin{lemma}\label{fast_fading_infinite}
\emph{As the  dataset size grows  ($N\rightarrow\infty$), }
%\begin{equation}
%\min_{\boldsymbol\mu_{\ell}}\underset{{i \in {\mathcal S}_{\ell}}} \sum d^2_p(\boldsymbol\Upsilon^{(i)}, \boldsymbol\mu^{T}_{\ell}).
%\end{equation}
\begin{equation}
\underset{{i \in {\mathcal C}_{\ell}}} \sum\emph{tr}\l\{ ({\bY^{(i)}})^{H}\boldsymbol\mu_{\ell}\boldsymbol\mu^{H}_{\ell}\bY^{(i)}  \r\} \! \longrightarrow \!\underset{{i \in {\mathcal C}_{\ell}}} \sum \l[N_t - d^2_p\l(\boldsymbol{\Upsilon}^{(i)}, \boldsymbol\mu_{\ell}\r)\r], \qquad \forall \ell .
\end{equation}
\end{lemma}
The proof is provided in Appendix~\ref{proof:fast_fading_infinite}.  Substituting the result in Lemma \ref{fast_fading_infinite} into \eqref{Eq:BookOptim} yields the following asymptotic form of the  M-step in \eqref{M_step_temp} in the case of high SNR and large dataset size: 
\begin{equation}\label{codeword_temp}
\hat {\boldsymbol{\mu}}_\ell =\arg\min_{\boldsymbol{\mu}_{\ell}\in \mathcal{G}} \sum_{i\in \mathcal{C}_\ell}d^2_p\l(\boldsymbol{\Upsilon}^{(i)}, \boldsymbol\mu_{\ell}\r), \quad \forall \ell.
\end{equation}
In this form, the M-step updates each codeword by computing the Grassmann centroid  of the cluster of Grassmann symbols associated with the codeword in the E-step in \eqref{E_step_temp}. 

\subsubsection{Asymptotic EM Algorithm}

Combining the results in \eqref{eq:DistanceAssign} and \eqref{codeword_temp}, in the case of a high SNR and a large dataset size, the asymptotic EM algorithm for detecting the Grassmann codebook $\mathcal{F}$  iterates between the following two steps: 
\begin{align}\label{eq:symbol detection}
(\text{Symbol detection})\quad  &\hat{\bX}^{(i)} =  \arg\min_{\hat{\boldsymbol\mu}_{\ell}\in \hat{\mathcal{F}}}  d^2_p\l(\boldsymbol\Upsilon^{(i)}, \hat{\boldsymbol\mu}_{\ell}\r), \quad \forall i,\\
(\text{Codeword optimization})\quad  &\hat{\boldsymbol\mu}_{\ell} = \arg\min_{\boldsymbol\mu_{\ell}\in \mathcal{G}}\sum_{i \in \mathcal{C}_\ell} d^2_p(\boldsymbol\Upsilon^{(i)}, \boldsymbol\mu_{\ell}), \quad \forall \ell\label{convertion to distance}.
\end{align}
This is exactly the  well-known Grassmann K-means algorithm, thereby relating the ML constellation detection to data clustering on the Grassmannian.

\vspace{-10pt}\section{Grassmann Constellation Detection by  Data Clustering}\label{Algorithm_constellation_detection}

In the preceding section, the ML constellation detection is shown to be asymptotically equivalent to Grassmann data clustering under a high SNR. In this section, building on this connection, several algorithms for Grassmann data clustering are briefly discussed  and applied to constellation detection. Furthermore, it is even possible to detect a Grassmann constellation without the knowledge of the constellation size, which is required by the previously considered EM algorithm for ML detection. 

\vspace{-10pt}\subsection{Data Clustering with a  Known Constellation Size}\label{kmean}
Consider the case that the constellation size,  $L =|\mathcal{F}| $,  is  known at the receiver. As derived in the preceding section, the Grassmann K-means algorithm for constellation detection iterates between two steps: 1) symbol detection in  \eqref{eq:symbol detection} and 2) codeword optimization in \eqref{convertion to distance} until convergence. An efficient implementation of the algorithm is proposed in \cite{turaga} and presented in Algorithm~\ref{Grassmann K-means} that replaces the current Procrustes distance with the geodesic distance as defined in \eqref{eq:distance}. This allows the step of codeword optimization in \eqref{convertion to distance} to be efficiently solved using the following algorithm of \emph{sample Karcher mean}. 

Considering  a  cluster of Grassmann symbols, say $\{i\in\mathcal{C}_\ell\}$, the sample Karcher mean, denoted as $\hat{\boldsymbol\mu}_\ell$,   can be defined as follows \cite{karcher}:
\begin{equation}
\hat{\boldsymbol\mu}_\ell = \arg \min_{\boldsymbol\mu_{\ell}\in {\cal{G}}}\frac{1}{N_\ell}\sum_{i\in \mathcal{C}_\ell}d_g^{2}\l(\boldsymbol\mu_{\ell},\boldsymbol\Upsilon^{(i)}\r)\label{eq:sample mean}. 
\end{equation}

\begin{algorithm} [tt]
	\caption{K-means Algorithm for Grassmann Constellation and Symbol Detection }
	\label{Grassmann K-means}
	\textbf{Input}: A block of Grassmann symbols  $\{\boldsymbol\Upsilon^{(i)}\}_{i=1}^{N}$  and the constellation size   $L$.\\
	\textbf{Output}: The estimated codewords  $\{\hat{\boldsymbol\mu}_{\ell}\}$ of   the  Grassmann constellation $\hat{\mathcal{F}}$.\\
	\textbf{Initialization}: Randomly choose $L$ symbols from  $\{\boldsymbol\Upsilon^{(i)}\}$ as the initial codewords.\\
	\textbf{Iterate}
	\begin{itemize}
	\item \textbf{Step 1 (Symbol Detection)}: Separate the symbols into $L$ clusters each is associated with  a single codeword. To this end,   assign  each Grassmann symbol, say $\boldsymbol\Upsilon^{(i)}$,  to the codeword  with the shortest geodesic distance, namely  $\hat{\bX}^{(i)} = \arg\min\limits_{\hat{\boldsymbol{\mu}}_{\ell}\in\mathcal{\hat{F}}}d^2_{g}(\boldsymbol\Upsilon^{(i)}, \hat{\boldsymbol{\mu}}_{\ell})$. 
	
		\item{\textbf{Step 2 (Codeword Optimization)}: For each symbol  cluster,  update the associated codeword  as the sample  Karcher mean of the cluster that is computed using Algorithm \ref{Sample Karcher Mean Algorithm}.}
		
	\end{itemize}
	\textbf{Until Convergence}	
\end{algorithm}

\begin{algorithm} [tt]
	\caption{ Algorithm of Sample Karcher Mean for Codeword Optimization}
	\label{Sample Karcher Mean Algorithm}
	\textbf{Input}: A block of Grassmann symbols $\{ \boldsymbol\Upsilon^{(i)} \}_{i=1}^M$.\\
	\textbf{Output}: The Karcher mean of the cluster, denoted as  ${\boldsymbol\mu}^{*}$.\\
	\textbf{Initialization}:  Set ${\boldsymbol\mu}^{*}$ as a randomly selected point from $\{ \boldsymbol\Upsilon^{(i)} \}$.\\
	\textbf{Iterate}
	\begin{itemize}
	\item{\textbf{Step 1}: Project the points in $ \{\boldsymbol\Upsilon^{(i)}\}$ onto the  tangent space  with ${\boldsymbol\mu_0}={\boldsymbol\mu}^{*}$ as the point of tangency by applying the logarithm mapping in \eqref{eq:log_orig}, i.e., $\bT^{(i)} = {\log}_{\boldsymbol\mu_0}( {\boldsymbol\Upsilon}^{(i)})$.
	}
	\item{\textbf{Step 2}: Calculate the mean direction $\bar{\bT}$ in the tangent space by averaging: $\bar{\bT} = \frac{1}{M}\sum_{i=1}^{M}{\bT^{(i)}}$.}
	\item{\textbf{Step 3}: Update the Karcher mean ${\boldsymbol\mu}^{*}$ by moving it in the direction of $\bar{\bT}$ via the exponential mapping in \eqref{eq:exp}:    ${\boldsymbol\mu}^{*} = {\exp}_{{\boldsymbol\mu}_0}(\tau\bar{\bT})$,  where the step size $\tau$ is typically set as $0.5$.}
	\end{itemize}
	\textbf{Until Convergence}.
\end{algorithm}

\noindent One can observe that the definition is equivalent to the derived codeword-optimization step in \eqref{convertion to distance} except for replacing the Procrustes distance with the geodesic distance. The algorithm of sample Karcher mean as presented in Algorithm \ref{Sample Karcher Mean Algorithm} solves  the optimization problem in \eqref{eq:sample mean} by gradient descend on the Grassmannian \cite{turaga,absil2009optimization}. The key idea of the algorithm is computing the descend direction  on the Grassmannian in a tangent Euclidean space exploiting exponential and logarithm mappings  between the two spaces [see \eqref{eq:exp} and \eqref{eq:log_orig}]. Last, it is worth mentioning that besides the Karcher mean, there exist other mean metrics such as   \emph{Procrustes mean} and related optimization algorithms \cite{chikuse}. As observed from simulation, the choices of the subspace distance metric  (e.g., geodesic versus Procrustes distances) and mean metrics of a cluster of Grassmann symbols (e.g., Karcher versus Procrustes  means) seem to have an insignificant effect on the performance of Grassmann constellation detection by data clustering. For this reason, the specific metric in a  particular part of analysis is selected for tractability without affecting the resultant general insights.

\vspace{-10pt}\subsection{Data Clustering with a  Unknown Constellation Size}\label{ms}

Consider the case that the constellation size,  $L =|\mathcal{F}| $,  is  unknown at the receiver. Without the knowledge, the K-means algorithm discussed in the last sub-section cannot be applied since it requires $L$ as the input. Specifically, the algorithm relies on randomly choosing $L$ Grassmann symbols as the centroids to generate  $L$ clusters. Alternatively, a standard  algorithm for \emph{connected-component identification} such as DFS~\cite{tarjan} can be applied to recognizing Grassmann symbol clusters by examining the pair-wise subspace distance against a pre-specified threshold denoted as $\gamma_0$. The main procedure of the DFS algorithm is summarized in Algorithm~\ref{DFS-Based Grassmannian Clustering Algorithm}. Note that a single calling of the DFS algorithm Algorithm outputs only one recognized cluster. As a result, repeatedly implementation of  DFS on the remaining  unlabelled symbols  is needed for resolving all clusters. 

Upon the completion of the DFS algorithm, the constellation size and the  estimated codewords can be computed as the number of clusters and their  sample Karcher means using \eqref{eq:sample mean}. Then  the received symbols are detected as their associated codewords.

\begin{algorithm} [tt]
	\caption{DFS-Based  Algorithm for Grassmann Symbol Clustering}
	\label{DFS-Based Grassmannian Clustering Algorithm}
	\textbf{Input}: The block of Grassmann symbols $\{ \boldsymbol\Upsilon^{(i)}\}_{i=1}^{N}$.\\
	\textbf{Output}: All $\{ \boldsymbol\Upsilon^{(i)}|\ \boldsymbol\Upsilon^{(i)} \neq \boldsymbol\Upsilon \}$ reachable from $\boldsymbol\Upsilon$ labeled as discovered.\\
	\textbf{Procedure} DFS(${\cal M},\boldsymbol\Upsilon$):
	\begin{itemize}
	\item{ Label $\boldsymbol\Upsilon$ as discovered.}
	\item{ \textbf{For all} $\{\boldsymbol\Upsilon^{'}\}$ in an adjacent set defined as ${\cal A}_{\boldsymbol\Upsilon} =\l\{ \boldsymbol\Upsilon^{(i)}|\ d_p\(  \boldsymbol\Upsilon^{(i)}, \boldsymbol\Upsilon\) \leq \gamma_0  \r\} \ \textbf{do}$}
	\item{ \textbf{If} $\boldsymbol\Upsilon^{'}$ is not labeled as discovered \textbf{then} recursively call DFS(${\cal M},\boldsymbol\Upsilon^{'}$).}
	\end{itemize}
	\label{Algo:DFS}
\end{algorithm}

\vspace{-10pt}\section{Performance of  Grassmann Constellation  Detection}\label{analysis: DFS}
Due to the difficulty in tractable analysis, there exists few theoretic result on  the performance of data clustering while prior work focuses on  algorithmic design (see e.g.,~\cite{turaga,tarjan}). In this section, we make an attempt to tackle the challenge by developing a framework for analyzing the performance of data clustering on the Grassmannian in the context of Grassmann constellation detection. In particular, by deriving the conditions of data forming well separable clusters, we can quantify the effects of various  system and algorithmic  parameters, ranging from the SNR to the connectivity threshold in the DFS algorithm, on the detection performance. 

\vspace{-10pt}\subsection{Approximate Signal Distribution}

\begin{figure} [tt]
\centering
\includegraphics[width = 8cm]{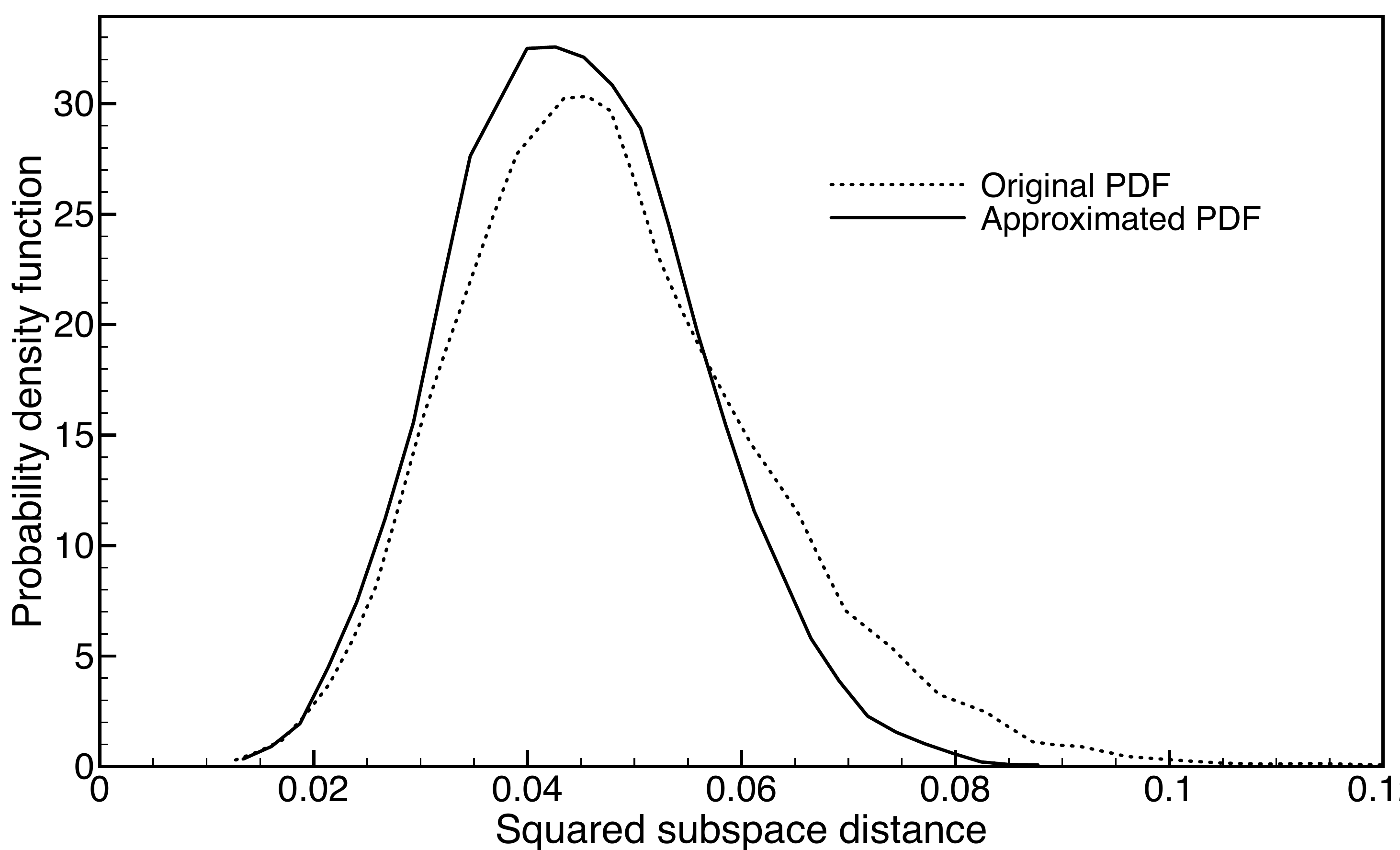}
\caption{Signal distribution approximation.}
\label{fig_pdf}
\end{figure}

A key step in the tractable analysis of Grassmann constellation detection is to approximate the distribution of received signals. Let $\textsf{span}(\bA)$ denote a basis spanning the column space of a matrix $\bA$. Then 
it follows from \eqref{decomposition_Y} that in the presence of noise, the received Grassmann  symbol $\boldsymbol\Upsilon^{(i)}$  is 
\begin{align}\label{Eq:RxSymb}
\boldsymbol\Upsilon^{(i)} = \mathsf{span}\l(\bX^{(i)} + \sqrt{\frac{N_t}{\rho T}} \bW^{(i)}\l[\begin{matrix} 
\lambda_1^{-1} &\cdots &0 \\ 
\vdots & \ddots & \vdots \\
0 & \cdots & \lambda_{N_t}^{-1} 
\end{matrix}\r]\r),
\end{align}
where $\bX^{(i)}$ is the transmitted (Grassmann) symbol and $\bW^{(i)}$ an i.i.d. Gaussian matrix representing noise. The  distribution of the random subspace distance of  $\boldsymbol\Upsilon^{(i)}$ from the centroid  $\bX^{(i)}$ determines the size of received signal cluster  centered at $\bX^{(i)}$. It is difficult to characterize the distribution due to the eclipse distribution of the noise process after scaling by the inverse channel singular values $\{\lambda_1^{-1} ,\cdots, \lambda_{N_t}^{-1}\}$. To overcome the difficulty, replacing all singular values in  \eqref{Eq:RxSymb} with the expectation of a typical one, denoted as $\bar{\lambda}$, yields a random
orthonormal matrix $\widetilde{\boldsymbol\Upsilon}^{(i)}$ defined as: 
\begin{align}\label{Eq:RxSymb:Approx}
\widetilde{\boldsymbol\Upsilon}^{(i)} = \mathsf{span}\l(\bX^{(i)} + \frac{1}{\bar{\lambda}}\sqrt{\frac{N_t}{\rho T}} \bW^{(i)}\r),
\end{align}
which results from $\bX^{(i)}$ perturbed by isotropic Gaussian noise. Then the distribution of the distance $d_p\l(\boldsymbol\Upsilon^{(i)}, \bX^{(i)}\r) $ is approximated by that of $d_p\l(\widetilde{\boldsymbol\Upsilon}^{(i)}, \bX^{(i)}\r)$: 
\begin{align}\label{approx_dist}
\text{(Approximate distance   distribution)}\quad d_p\l(\boldsymbol\Upsilon^{(i)}, \bX^{(i)}\r)  \overset{\text{d}}{\approx}   d_p\l(\widetilde{\boldsymbol\Upsilon}^{(i)}, \bX^{(i)}\r),
\end{align}
where $\overset{\text{d}}{\approx}$ represents approximation in  distribution. 
%The approximation is sufficiently accurate as shown in Fig.~\ref{fig_pdf}. 
\begin{remark}\emph{(Accurate  distance-distribution  approximation).
The approximation   in \eqref{approx_dist} is accurate in the case that  the transmit antennas are far outnumbered by  receive ones, i.e., $N_r \gg N_t$, and the resultant  large spatial diversity gain makes the channel matrix well conditioned with $\lambda_1 \approx \lambda_2 \cdots \approx \lambda_{N_r}$. Furthermore, empirical results with typical setting $N_t =2 , N_r = 10$ is provided in Fig.~\ref{fig_pdf} to further support the statement.}
\end{remark}

For convenience, given a codeword $\boldsymbol\mu_{\ell}$, denote  $d_p\l(\boldsymbol\Upsilon^{(i)}, \boldsymbol\mu_{\ell}\r)$ and $d_p\l(\widetilde{\boldsymbol\Upsilon}^{(i)}, \boldsymbol\mu_{\ell}\r)$ as $d_{\ell}^{(i)}$ and $\tilde{d}_{\ell}^{(i)}$, respectively. Unlike $d_{\ell}^{(i)}$, the distribution of $\tilde{d}_{\ell}^{(i)}$ is independent of the direction from $\boldsymbol\mu_{\ell}$ to $\widetilde{\boldsymbol\Upsilon}^{(i)}$ due to   the isotropicity of noise in \eqref{Eq:RxSymb:Approx}. As a result, the distribution of $\tilde{d}_{\ell}^{(i)}$,  which approximates   that of the desired  r.v. $d_{\ell}^{(i)}$, can be characterized mathematically. To this end, a useful result is provided. 
\begin{lemma}[\cite{chikuse}] \label{Lem:Chikuse}\emph{Let $\boldsymbol{\Upsilon} = \mathsf{span}(\boldsymbol{\mu} + \bA)$ with $\boldsymbol{\mu} \in\mathbb{O}^{M\times N}$ and $\bA$ an $M\times N$ matrix having i.i.d. $\mathcal{CN}(0, \sigma^2)$ elements. Then given $\boldsymbol{\mu}$ and as $\sigma^2 \rightarrow 0$,  the distance $d_p(\boldsymbol{\Upsilon}, \boldsymbol{\mu})$ has the following distribution: 
\begin{align}\label{Chi-squared}
\l[d_p(\boldsymbol{\Upsilon}, \boldsymbol{\mu}) \r]^2 \sim \sigma^2 {\mathcal X}^{2}_{D}, 
\end{align}
where $D  = 2N(M-N)$ and ${\mathcal X}^{2}_{D}$ represents a Chi-squared r.v. with $D$ degrees of freedom.
}
\end{lemma}
The distance $d^{(i)}_{\ell} $ defined earlier represents  the random deviation of a received symbol from the corresponding transmitted symbol. Using Lemma~\ref{Lem:Chikuse}, its  distribution is characterized as follows.

\begin{lemma}\label{lemma: coverage of a typical cluster}
\emph{Consider an arbitrary Grassmann codeword $\boldsymbol\mu_{\ell}$ and the approximation in~\eqref{approx_dist}. In the high-SNR regime ($\rho\rightarrow \infty$), $d_\ell^{(i)}\overset{\text{d}}{\approx}\tilde{d}_\ell^{(i)}$ with the distribution of $\tilde{d}_\ell^{(i)}$ given as 
\begin{align}
 \text{Pr}\(\tilde{d}^{(i)}_{\ell} \geq r\) &= \frac{1}{\Gamma(\frac{D}{2})} \Gamma\( \frac{D}{2}, \frac{\rho T \bar{\lambda}^{2} {r}^{2}}{2N_t} \), \qquad \forall i \in {\mathcal C}_{\ell}\\
& = \frac{\( \frac{\rho T \bar{\lambda}^{2} {r}^{2}}{2N_t}  \)^{\frac{D}{2}-1}}{\Gamma(\frac{D}{2})}\exp \( -\frac{\rho T \bar{\lambda}^{2} {r}^{2}}{2N_t}\)\( 1 + o\( \frac{1}{\rho}\)  \).
\end{align}
with $r \geq 0$ and the upper incomplete Gamma function $\Gamma(D, x) = \int_{x}^{\infty} t^{D-1}e^{-t} dt$. 
}
\end{lemma}
One can observe from the result that $\Pr(d_\ell^{(i)} \geq r)$ decays  exponentially as the SNR $\rho$ grows. This suggests that at a high SNR, received symbols tend to cluster around their corresponding transmitted codewords  and the clusters shrink rapidly as the SNR grows. This makes  them well separated, facilitating constellation detection using  a clustering algorithm. This insight is rigorously studied in the following sub-sections building on the approximation in~\eqref{approx_dist} and distance distribution in Lemma~\ref{lemma: coverage of a typical cluster}.

\vspace{-10pt}\subsection{Constellation Detection with a Known Size}

Considering the case that the receiver has prior knowledge of the  constellation size $L$ such that the the K-means algorithm in Algorithm~\ref{Grassmann K-means} can be applied to constellation detection. For the algorithm to be effective, the received symbols should form well separated clusters on the Grassmannian. In this section, the conditions for forming clusters are derived and then applied to study the effects of system parameters  on the algorithmic performance. 

%\footnote{XXX: Sometimes I use "constellation points" instead of "codewords" or "constellation" when "codebook" should be used. Please check and change to "codewords" and "codebook"}
First, a metric, called \emph{separability probability}, is defined to measure the level of clustering of the received symbols. To begin with, using the codewords $\{\boldsymbol \mu_\ell\}$ in $\mathcal{F}$ as centroids  and applying the nearest-neighbour rule, the Grassmannian $\mathcal{G}_{N_t, T}$ can be partitioned into $L$ \emph{Voronoi cells}. The cell with the centroid  $\boldsymbol \mu_\ell$ is  denoted as $\mathcal{V}(\boldsymbol \mu_\ell)$ and  defined  as 
\begin{equation}
\mathcal{V}(\boldsymbol \mu_\ell) =\l\{\boldsymbol{\Upsilon} \in \mathcal{G}_{N_t, T}\mid d_p(\boldsymbol{\Upsilon}, \boldsymbol \mu_\ell) <  d_p(\boldsymbol{\Upsilon}, \boldsymbol \mu_m)    \ \forall\  m \neq \ell \r\}. 
\end{equation}
Intuitively, the received symbol clusters are separable if each of them is  contained mostly within the \emph{correct} Voronoi cell, namely the one   having the  corresponding transmitted codeword  as the  centroid. Then an effective  initiation of the K-mean algorithm (see Algorithm~\ref{Grassmann K-means}),   namely the $L$ initial centroids are all within different Voronoi cells, can lead to convergence to their centroids or equivalently the correct detection of the  constellation.  Inspired by  this fact, we define the \emph{separability probability}  as the probability that a received Grassmann symbol lies in the correct  Voronoi cell. Then a larger  separability probability corresponds to a higher level of  separability of the received symbol clusters and hence  better performance of constellation detection, and vice versa. The mathematical definition of the metric is given below. 

\begin{definition}\emph{(K-means Separability Probability).   Let $\bX$ denote a typical transmitted symbol and $\boldsymbol{\Upsilon}$ the corresponding received symbol. The \emph{separability probability}, denoted as $p_{\sf sep}$, is defined as 
\begin{equation}
p_{\sf sep} = \frac{1}{L}\sum_{\ell =1}^L \Pr\l(\boldsymbol{\Upsilon} \in \mathcal{V}(\boldsymbol{\mu}_\ell)\mid \bX = \boldsymbol{\mu}_\ell \r). 
\end{equation}
}
\end{definition}

Though direct analysis of $p_{\sf sep}$ is difficult, a tractable lower bound can be obtained as follows. For the codebook $\mathcal{F}$, with the minimum codeword pairwise distance $d_{\min}$ defined in~\eqref{min_distance}. The optimal codebook design by packing in~\eqref{Eq:Packing} attempts to maximize $d_{\min}$. It is well known in the literature of Grassmannian packing that $d_{\min}$ can be bounded as (see e.g., \cite{barg2002bounds})
\begin{equation}\label{bound_dmin}
d^2_{\min} \geq 4N_t\(\frac{1}{L}\)^{\frac{1}{TN_t}}. 
\end{equation}
Given $d_{\min}$, a sufficient condition for a cluster of received symbols, say those with the indices $\mathcal{C}_\ell$,  originating from the same codeword, say $\boldsymbol{\mu}_\ell$, to be contained within the correct Voronoi cell is: 
\begin{equation}
\max_{i \in {\mathcal C}_{\ell}} d_p(\boldsymbol\Upsilon^{(i)},\boldsymbol\mu_{\ell}) \leq \frac{d_{\min}}{2}. \nn
\end{equation}
Then jointly considering   the  sufficient conditions for all clusters of symbols leads to 
\begin{align}
p_{\sf sep} \geq \text{Pr}\( \bigcap_{\ell =1}^{L} \max_{i \in {\mathcal C}_{\ell}} d_p(\boldsymbol\Upsilon^{(i)},\boldsymbol\mu_{\ell}) \leq \frac{d_{\min}}{2}\).
\end{align}
Combining this result and that in Lemma~\ref{lemma: coverage of a typical cluster} gives the following main result of the sub-section. 

\begin{theorem}[K-means Separability Probability] \label{thm_kmean}
\emph{Consider Grassmann constellation detection using the K-means algorithm. In the  high SNR regime ($\rho\rightarrow \infty$), the separability probability satisfies} 
\begin{align}\label{eq:L-cluster separation kmeans}
p_{\sf sep} & \geq  \l [ \frac{1}{\Gamma(\frac{D}{2})} \gamma\( \frac{D}{2}, \frac{\rho T \bar{\lambda}^{2} d^2_{\min}}{8N_t} \)  \r]^{N} \\
& = 1-Ne^{-\frac{\rho T \bar{\lambda}^{2} d^2_{\min}}{8N_t}}G_m(\rho) + O(e^{-2\rho}),\qquad \rho \rightarrow \infty\label{inter-cluster-exp-kmean}, 
\end{align}
\emph{where $G_m(\rho)$ is a polynomial function of $\rho$ defined as $G_m(\rho) = \sum_{m=0}^{\frac{D}{2}-1}\frac{(T\bar{\lambda}^2d^2_{\min})^{m}}{m! (8N_t)^{m}}{\rho}^{m}$ and $\gamma$ denotes the lower incomplete Gamma function defined as $\gamma(D,x) = \int_{0}^{x} t^{D-1}e^{-t} dt$. }
\end{theorem}

By measuring the performance of constellation detection by the separability probability, the effects of two parameters, the SNR and dataset size, on the performance can be inferred from the result in Theorem~\ref{thm_kmean} as described below. 

\begin{itemize}

\item {\bf Effect of SNR}: One can observe from \eqref{inter-cluster-exp-kmean} that $p_{\sf sep}$ converges to one  exponentially fast as $\rho$ grows.  Intuitively, in the  high SNR regime, the received symbols form highly compact clusters on the Grassmannian. This enhances   the pairwise differentiability of the clusters and leads to accurate constellation detection.

\item {\bf Effect of Dataset Size}: According to \eqref{inter-cluster-exp-kmean}, in the high SNR regime, the separability probability may decay linearly with the dataset size $N$ as confirmed by simulation. The reason is that as the dataset size grows, it is more likely that there exist symbols having large distances from the centroids of their correct Voronoi cells. As a result, the separation gaps between clusters narrow or they even overlap, increasing the difficulty in  accurate clustering and thereby degrading the detection performance.

\item {\bf Dataset-SNR Tradeoff}: Based on \eqref{inter-cluster-exp-kmean}, the lower bound on $p_{\sf sep}$ can be written in a simple form to reflect the tradeoff between the SNR and dataset size:
\begin{equation}
p_{\sf sep} \approx  1 - e^{\log N -c \rho }, \qquad \rho \rightarrow \infty,
\end{equation}
with $c$ being a constant. One can infer from the result that under a constraint on the separability probability, as $N$ grows, the SNR should scale up linearly with  $\log N$. 

\item {\bf Effect of Constellation Size}: The dependency of $p_{\sf sep}$ on  $d^2_{\min}$ in \eqref{inter-cluster-exp-kmean} can be further translated  to that on  $L$. Specifically, by substituting \eqref{bound_dmin} to \eqref{inter-cluster-exp-kmean}, 
\begin{align}\label{psep_vs_L}
p_{\sf sep} \approx  1 - a_0 e^{-b_0   \rho L^{-\frac{1}{TN_t} }}, \qquad \rho \rightarrow \infty,
\end{align}
where $a_0$ and $b_0$ are constants. It can be clearly seen that $p_{\sf sep}$ monotonically decreases with respect to $L$. This aligns with our intuition that packing more constellation points (codewords) on a fixed Grassmann manifold will decrease $d_{\min}$, thus making different clusters harder to be distinguished.
Furthermore, one can infer from the result that given a target separability probability, as $L$ grows, the SNR should approximately scale up linearly with  $ L^{\frac{1}{TN_t}}$.

\end{itemize}

\vspace{-10pt}\subsection{Constellation Detection with an Unknown Size}\label{DFS detection}
Considering the case that  the  constellation size $L$ is unknown at the receiver and  the DFS algorithm in Algorithm~\ref{Algo:DFS} is  applied to constellation detection.  The algorithm is based on a different principle from that   of the K-means algorithm in the preceding  case. While K-means relies on iterative  centroid computation  and clustering, the DFS attempts to connect neighbouring symbols to form clusters by applying a distance threshold $\gamma_0$ (see Algorithm~\ref{Algo:DFS}), called the DFS threshold. Consequently, two factors of the dataset distribution affect the DFS performance. One is the separability of symbol clusters as for the K-means algorithm, which is measured by the separability probability. By slight abuse of notation, the metric for the DFS is also denoted as $p_{\mathsf{sep}}$.   The other is the connectivity within each single cluster, which is unique for the DFS. A metric, called \emph{connectivity probability} and denoted as $p_{\mathsf{con}}$, is defined in the sequel to measure the intra-cluster connectivity of the received dataset. Given the metrics, the effectiveness of constellation detection by the DFS can be ensured by applying constraints on their values: 
\begin{equation}\label{Eq:DFS:Constraints}
p_{\sf sep} \geq 1-\epsilon, \quad p_{\sf con} \geq 1-\delta,
\end{equation}
where $0 < \epsilon, \delta < 1$. In the sequel, $p_{\mathsf{sep}}$ and $p_{\mathsf{con}}$ are analyzed separately and  the results are then combined to quantity the effects the parameters of the system and algorithm on the detection performance.

\subsubsection{Inter-cluster Separation} For the DFS, the separation between two clusters of Grassmann symbols specified by the index sets $\mathcal{C}_m$ and $\mathcal{C}_\ell$ can be measured by the minimum pairwise distance, referred to as the \emph{inter-cluster distance} and  defined mathematically as 
\begin{equation}\label{eq: inter-cluster distance}
d_{\mathsf{clu}}({\mathcal C}_m,{\mathcal C}_\ell) = \min_{ i \in {\mathcal C}_{m}, j \in {\mathcal C}_{\ell}}d_p(\boldsymbol\Upsilon^{(i)},\boldsymbol\Upsilon^{(j)}).
\end{equation}
The two clusters can be separated by the DFS when their distance exceeds the DFS threshold $\gamma_0$. Based on this fact, the separability probability for the DFS can be defined as follows. 
\begin{definition}\emph{(DFS Separability Probability).  For constellation detection using the DSF algorithm, the \emph{separability probability} $p_{\sf sep}$ is defined as 
\begin{equation}\label{eq_inter_exact}
p_{\sf sep} = \text{Pr}\l( \min_{m \neq \ell} d_{\mathsf{clu}}\l({\mathcal C}_{m},{\mathcal C}_{\ell}\r) > \gamma_0\r).
\end{equation}
}
\end{definition}

Though the direct analysis of $p_{\sf sep}$ is difficult, a lower bound can be derived by designing a sufficient condition for cluster separation. Specifically, given the codebook $\mathcal{F}$ with $d_{\min}$, the symbol clusters are separable in terms of the criterion in \eqref{eq_inter_exact} if all received symbols deviate from their transmitted  codewords no more than a distance of $ \frac{d_{\min}-\gamma_0}{2}$ (see Fig.~\ref{fig_cluster_bin}). Therefore, $p_{\sf sep}$ can be lower bounded as
\begin{equation}\label{pair_separation}
p_{\sf sep}  \geq \prod_{\ell=1}^L \text{Pr}\l( \max \limits_{i \in {\mathcal C}_\ell} d_p\l(\boldsymbol\Upsilon^{(i)},\boldsymbol\mu_\ell\r) \leq \frac{d_{\min}-\gamma_0}{2}\r).
\end{equation}

Following the same procedure for deriving Theorem~\ref{thm_kmean}, we obtain the following corollary. 

\begin{corollary}[DFS Separability Probability] \label{thm_dfs}
\emph{Consider Grassmann constellation detection using the DFS algorithm. In the  high SNR regime ($\rho\rightarrow\infty$), the separability probability satisfies} 
\begin{align}\label{eq:L-cluster separation}
p_{\sf sep} ({\rho, L, N, \gamma_0}) & \geq  \l[ \frac{1}{\Gamma(\frac{D}{2})} \gamma\( \frac{D}{2}, \frac{\rho T \bar{\lambda}^{2} {(d_{\min}-\gamma_0)}^{2}}{8N_t} \)  \r]^{N}\\
& = 1-Ne^{-\frac{\rho T \bar{\lambda}^{2} {(d_{\min}-\gamma_0)}^{2}}{8N_t}}C_m(\rho)+O(e^{-2\rho})\label{inter-cluster-exp}, 
\end{align}
\emph{where  $C_m(\rho)$ is a polynomial function of $\rho$ defined as  $C_m(\rho) = \sum_{m=0}^{\frac{D}{2}-1}\frac{(T\bar{\lambda}^2(d_{\min}-\gamma_0)^2)^{m}}{m! (8N_t)^{m}}{\rho}^{m}$. }
\end{corollary}
The effects of the parameters including SNR, dataset size and constellation size are similar to their K-means counterparts discussed in the preceding sub-section. A remark is given below on the effect of the DFS threshold $\gamma_0$. 
\begin{remark}[Effect of DFS Threshold] \emph{Choosing a too small  value of the threshold  $\gamma_0$ leads to the failure of connecting points within a same cluster and thereby causes  it to be split into multiple clusters. On the other hand, if $\gamma_0$ is too large, multiple clusters may be connected into a single one. Both cases lead to incorrect constellation detection. Thus $\gamma_0$ should be optimized in practice to balance inter-cluster separability  and intra-cluster connectivity.}
\end{remark}

\subsubsection{Intra-cluster Connectivity}The analysis of intra-cluster connectivity is much more challenging than that of inter-cluster separation. In the context of DFS, two points  on the Grassmannian are \emph{neighbours } if their subspace distance is shorter than $\gamma_0$. A path is a sequence of points where every pair of adjacent points are neighbours. Then two points  are \emph{connected} if there exists a path connecting them. Based on this definition, the direct analysis of connectivity probability is intractable. Inspired by the analysis in the classic area of network connectivity (see e.g,~\cite{gupta1999critical}), we develop a  geometric technique for deriving a lower bound on the metric and its principle is described as follows. 

\begin{figure} [tt]
\centering
\includegraphics[width = 11.5cm]{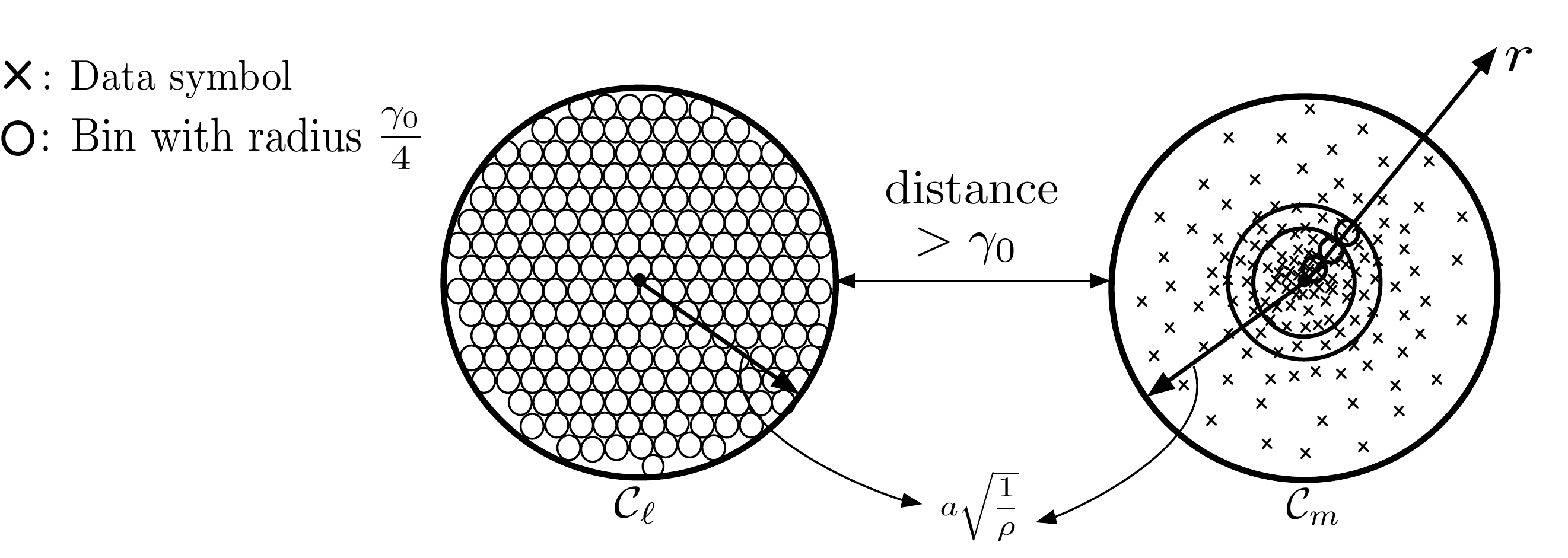}
\caption{Illustration of pairwise clusters.}
\label{fig_cluster_bin}
\end{figure}

\noindent {\bf Principle of Connectivity Analysis:} Consider a cluster of points (symbols) on the Grassmannian that are bounded by a disk. The disk is then packed  by uniform bins (small disks) each with a diameter $\frac{\gamma_0}{2}$ as illustrated in Fig.~\ref{fig_cluster_bin}. As a result, a sufficient condition for all points in the cluster being connected is that all bins are non-empty, namely that  each bin contains at least one point. The probability of this event can be derived in closed form that  lower bounds the connectivity probability. 

Based on the principle, the specific mathematical technique is developed and the desired result obtained as follows. 
First, for ease of exposition, consider the (intra-cluster) \emph{disconnect probability} defined as $p_{\mathsf{dis}} = 1- p_{\mathsf{con}}$. Consider the symbols cluster corresponding to the transmitted codeword $\boldsymbol{\mu}_\ell$.  Let $p_{\mathsf{dis}}(N_\ell)$ denote the   disconnect probability for the cluster conditioned the cluster size $N_{\ell}$. Then  $p_{\mathsf{dis}} = \mathbb{E}[p_{\mathsf{dis}}(N_\ell) ]$. Since the $L$  codewords have equal probabilities to be transmitted, $N_\ell$ follows the binomial distribution with parameters $N$ and $1/L$, i.e. $N_{\ell} \sim B\(N, \frac{1}{L}\)$.

Next, consider a cluster of symbols originating from the same transmitted codeword $\boldsymbol{\mu}$. A disk with the centroid $\boldsymbol{\mu}$ and a radius $r$ is defined on the Grassmannian as $\mathcal{B}(\boldsymbol{\mu}, r) = \{\boldsymbol{\Phi}\in \mathcal{G}\mid d_p(\boldsymbol{\Phi}, \boldsymbol{\mu})\leq r\}$. It is known in the literature that  in the presence of Gaussian noise, the received symbols with the  transmitted codeword $\boldsymbol{\mu}$  lie with high probability  in a disk $\mathcal{B}(\boldsymbol{\mu}, r)$, whose  radius $r$ is proportional to the standard deviation of noise  or equivalently  proportional to $\frac{1}{\sqrt{\rho}}$ with $\rho$ being the SNR \cite{gohary2009,zheng2002communication}. Therefore, the disk radius can be chosen as $\frac{a}{\sqrt{\rho}}$ with $a$ being a constant (see Fig.~\ref{fig_cluster_bin}). The constant can be appropriately chosen such that a symbol lies within the disk with probability no smaller than e.g., $(1-\frac{\epsilon}{N})$, which, as implied by~\eqref{inter-cluster-exp}, is sufficient for satisfying the separability constraint in \eqref{Eq:DFS:Constraints}. 
\begin{assumption}\label{DFS_Assumption}
\emph{The  dataset size $N$ is sufficiently large such that the points within each disk are dense. Then the required DFS threshold $\gamma_0$ for connecting the points within a  disk is much smaller  than its radius:   $\gamma_0 \ll \frac{a}{\sqrt{\rho}}$.
}
\end{assumption}
Based on the assumption, the disk  can be packed  with small disks each with the diameter $\frac{\gamma_0}{2}$, called \emph{bins}, as illustrated in Fig.~\ref{fig_cluster_bin}. Each of the bins  thus is placed contacted with at least one another bin. The cluster of symbols can be treated as i.i.d. random points. A bin is nonempty if it contains at least one point. In the event that all bins are nonempty, all points are guaranteed to be connected regardless of if they are inside or outside bins. Therefore, given that the number of points in the cluster is $N_\ell$, the corresponding disconnect probability can be lower bounded as 
\begin{align}\label{Eq:Ineq}
p_{\mathsf{dis}}(N_\ell) \leq \text{Pr}(\ \exists \ \text{one empty bin}| N_{\ell} ).
\end{align}
Note that the number of bins in the disk is $M = \eta_{_D} \(\frac{\frac{a}{\sqrt{\rho}}} {\frac{\gamma_0}{4}}\)^{D}$ where $\eta_{_D}$ represents the fraction of the disk area covered by bins which is a constant given the space dimensions of $D$. Define an indicator  function $\mathbb{I}(\mathcal{A}_{i}) = 1$ if the $i$th bin is empty, and  $\mathbb{I}(\mathcal{A}_{i}) = 0$ otherwise. The  inequality in \eqref{Eq:Ineq} can be rewritten by
\begin{align}\label{Eq:Ineq:1} 
p_{\mathsf{dis}}(N_\ell) \leq \text{Pr}\( \sum_{i=1}^{M}\mathbb{I}(\mathcal{A}_{i}) \geq 1 |  N_{\ell} \),
\end{align}
By applying  Markov inequality, 
\begin{equation}
p_{\mathsf{dis}}(N_\ell) \leq \mathbb{E}\( \sum_{i=1}^{M}\mathbb{I}(\mathcal{A}_{i})| N_{\ell} \) = \sum_{i=1}^{M}(1-p_i)^{N_{\ell}},
\end{equation}
where $p_i$ denotes the probability that a typical point falls into the $i$th bin. Define $p_{\min} = \min \limits_{i}p_i$. It follows from \eqref{Eq:Ineq:1} that 
\begin{align} \label{condi_prob_specific}
p_{\mathsf{dis}}(N_\ell) \leq M(1-p_{\min})^{N_{\ell}}.
\end{align}
By invoking  the Binomial  distribution of $N_{\ell}$, 
\begin{align}
p_{\mathsf{dis}} = \mathbb{E}[p_{\mathsf{dis}}(N_\ell)]  &\leq  M\(1-\frac{p_{\min}}{L}\)^{N}\nn. 
\end{align}
Then the   result below  follows.

\begin{lemma}\label{intra_upper_bound}
\emph{In the high SNR regime, the disconnect  probability satisfies: $p_{\mathsf{dis}}  \leq Me^{-\frac{p_{\min}}{L}N}$.}
\end{lemma}

Next, to obtain a concrete upper bound on $p_{\mathsf{dis}}$, an expression is derived for $p_{\min}$ as follows. In the presence of isotropic noise, the probability that a receive symbol $\boldsymbol{\Upsilon}$ originating from a codeword $\boldsymbol{\mu}$  falls into a bin $\mathcal{B}(\boldsymbol{\Phi},\frac{\gamma_0}{4})$ depends on the distance $d_p(\boldsymbol{\Phi}, \boldsymbol{\mu})$ as well as the bin volume, denoted as $\textrm{Vol}_{\text{bin}}$, but is independent of the direction from $\boldsymbol{\mu}$ to $\boldsymbol{\Phi}$. Define a ring with the center $\boldsymbol{\mu}$, width $\frac{\gamma_0}{2}$, and radius $r$ as $\mathcal{R}(\boldsymbol{\mu}, r) = \{\boldsymbol{\Phi}\in \mathcal{G}\mid r - \frac{\gamma_0}{2}\leq d_p(\boldsymbol{\Phi}, \boldsymbol{\mu})\leq r\}$ which is illustrated in Fig.~\ref{fig_cluster_bin}. Then the symbol $\boldsymbol{\Upsilon}$ falls with equal probabilities into the bins lying in  a same ring $\mathcal{R}(\boldsymbol{\mu}, r)$. Let the probability be denoted as $p(r)$ and the volume of the ring as $\textrm{Vol}_{\text{rin}}(r)$. Then 
\begin{align}\label{eq_location}
p(r) &=  \frac{\eta^{-1}_{_D}\textrm{Vol}_{\text{bin}}(r)}{\textrm{Vol}_{\text{rin}}(r)} \times \text{Pr}\(r-\frac{\gamma_0}{2} \leq d_p(\boldsymbol{\Upsilon}, \boldsymbol{\mu}) \leq r\)\nonumber\\
& \overset{(a)}{=} \frac{\eta^{-1}_{_D}(\frac{\gamma_0}{4})^{D}}{r^{D}-(r-\frac{\gamma_0}{2})^{D}} \times \frac{1}{\Gamma(\frac{D}{2})}\l\{ \Gamma\(\frac{D}{2},\frac{\rho T{\bar\lambda^2}(r-\frac{\gamma_0}{2})^2}{2N_t}\)\,-\,\Gamma\(\frac{D}{2},\frac{\rho T{\bar\lambda}^2r^2}{2N_t}\)\r\}, \ (r \geq \frac{\gamma_0}{2}),
\end{align}
where $D = 2N_t(T-N_t)$ is the dimensions and $(a)$ applies the distance distribution in~\eqref{Chi-squared}. Given $p(r)$, $p_{\min}$ can be equivalently written as $p_{\min} = \min_{\frac{\gamma_0}{2}\leq r\leq \frac{a}{\sqrt{\rho}}}p(r)$. By analyzing the   derivative of $p(r)$, it is straightforward to prove that the function is monotonically decreasing in the range of $r \geq \frac{\gamma_0}{2}$ (see Appendix \ref{proof: monoton_property}), leading to the following result. 

\begin{lemma}\label{Lemma: monoton_property}
\emph{If  the disk radius $\frac{a}{\sqrt{\rho}} \geq \frac{\gamma_0}{2}$, $p_{\min} = p\l(\frac{a}{\sqrt{\rho}}\r)$ with $p(r)$ given in \eqref{eq_location}.}
\end{lemma}
%\proof
%See Appendix \ref{proof: monoton_property}.
%\endproof

 The above lemma shows that the bin with $p_{\min}$ locates at the boundary of the disk. Under Assumption~\ref{DFS_Assumption} and using  \eqref{eq_location} and Lemma~\ref{Lemma: monoton_property}, a simplified  asymptotic expression for $p_{\min}$ can be derived as: 
 \begin{align}\label{p_min_approx}
p_{\min} = \frac{\eta^{-1}_{_D}2^{-\frac{5D}{2}+1}}{D\Gamma\(\frac{D}{2}\)}\(\frac{T{\bar\lambda}^2}{N_t}\)^{\frac{D}{2}}{\gamma_0}^{D}\rho^{\frac{D}{2}}e^{-\frac{a^2T{\bar\lambda}^2}{2N_t}} + o({\gamma_0}^D{\rho}^{\frac{D}{2}}).
\end{align}
The derivation details can be found in Appendix~\ref{proof: p_min_calculation}. Finally, substituting \eqref{p_min_approx} and $M = \eta_{_D} \(\frac{4a}{\gamma_0\sqrt{\rho}}\)^{D}$ into the result in Lemma~\ref{intra_upper_bound}, we can derive a lower bound of the success probability of intra-cluster connectivity, which is presented as follows.

\begin{theorem}[DFS Connectivity Probability] \label{Thm:DFS:Connect}
\emph{In the high SNR regime, the connectivity probability satisfies 
\begin{align}\label{lower_final}
p_{\sf con}   \geq 1-\eta_{_D}(4a)^{D}{\gamma_0}^{-D}\rho^{-\frac{D}{2}}e^{-c_0{\gamma_0}^{D}\rho^{\frac{D}{2}}\frac{N}{L}},
\end{align}
where $c_0 = \frac{\eta^{-1}_{_D}2^{-\frac{5D}{2}+1}}{D\Gamma\(\frac{D}{2}\)}\(\frac{T{\bar\lambda}^2}{N_t}\)^{\frac{D}{2}}e^{-\frac{a^2T{\bar\lambda}^2}{2N_t}}$ is a constant and $N/L$ denotes the expected number of received symbols in each cluster.}
\end{theorem}

\subsubsection{Effects of Parameters on Detection Performance}
Comparing the results in Corollary~\ref{thm_dfs} and Theorem~\ref{Thm:DFS:Connect}, we obtain the following insights into the effects on various parameters on the constellation detection performance. 

\begin{itemize}

\item \textbf{Effect of SNR}: One can observe from \eqref{inter-cluster-exp} and \eqref{lower_final} that both $p_{\sf sep}$ and $p_{\sf con}$ converge \emph{exponentially} to one as $\rho$ grows. A higher SNR makes the  dataset distributed in  more concentrated clusters centered at the codewords, improving their separability and connectivity in terms of $p_{\sf sep}$ and  $p_{\sf con}$, respectively.

\item \textbf{Effect of Dataset Size}: Unlike the SNR, the effect of increasing $N$ is double-sided. On one hand, $\eqref{lower_final}$ suggests that the intra-cluster connectivity improves exponentially with growing  $N$ due to the increasing point-density of each   cluster. On the other hand, \eqref{inter-cluster-exp} shows that the separability between clusters may decrease exponentially as $N$ increases. This is because that increasing $N$ may shorten the \emph{inter-cluster distance} defined in~\eqref{eq: inter-cluster distance} due to the more likely  existence of ``outliers" and the resultant growth of cluster radius.

\item \textbf{Effect of Constellation Size}: Last,  a larger constellation size $L$ reduces  both $p_{\sf sep}$ and $p_{\sf con}$ and makes it harder to perform accurate detection by DFS algorithm. Specifically, one can observe from  \eqref{eq:L-cluster separation} that the separability of different clusters reduces  as $L$ increases. This is aligned  with our intuition that packing more constellations points on a fixed Grassmann manifold reduces $d_{\min}$, thereby increasing the difficulty of clustering in the presence of noise. Moreover, given the dataset size  $N$, as suggested by \eqref{lower_final}, a smaller $L$ benefits intra-cluster connectivity since each cluster is expected to comprise more points (the expected number of points is given by ${N}/{L}$), thus denser clusters are formed.
\end{itemize}

\begin{figure} [tt]
\centering
\includegraphics[width = 7cm]{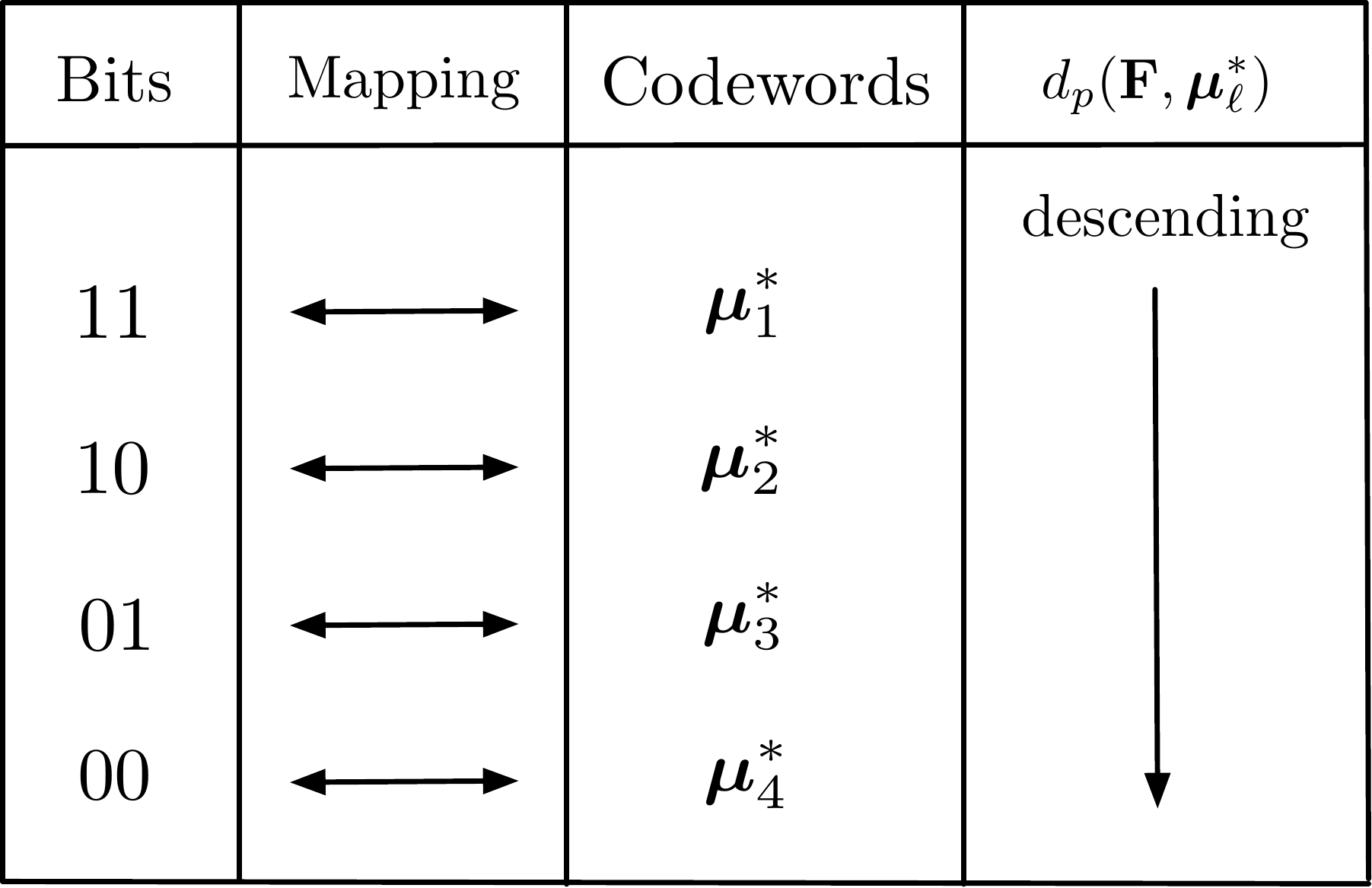}
\caption{Illustration of bit-symbol-mapping for constellation size of four.}
\label{Bit_symbol_mapping}
\end{figure}

\section{Constellation Embedded Bit-Symbol Mapping}\label{bit-symbol-error mapping}

Given the inferred constellation codewords, the information retrieval process contains two substeps: 1) associate the observed data to the closest constellation codeword in terms of their distance; 2) map the codeword to corresponding bit sequence according to a pre-defined mapping rule. In this sub-section, we aim to propose an intelligent mechanism for resolving the mapping between the constellation codewords and the embedded  information bits without compromising the spectrum efficiency. Specifically, the novel scheme we proposed encodes the mapping information to the subspace distance between the transmit codewords and a well-devised orthonormal reference point such as a truncated Fourier matrix, denoted by $\bF$. Concretely, the transmit codewords are one-to-one mapped to a set of information bits following a pre-defined order determined by their subspace distances to the selected reference point  (see Fig. \ref{Bit_symbol_mapping}). The order that encodes the mapping information can be accurately recovered at the receiver since the subspace distances between codewords and $\bF$ are invariant to the channel rotation. Note that the reference point should be carefully selected to ensure the subspace distance differentiation to each codeword. To this end, two candidate schemes are proposed: 1) fix a reference point  first and select from a set of packing based codebooks the one having the most differentiation of subspace distances; 2) fix a packing based codebook first and then choose the optimal  in terms of subspace distance differentiation. The advantage of the scheme 2) over 1) is that codewords only need to be generated once, but at the additional expense of reference point transmission. Note that communicating the reference point can incur overhead (coding and high power) due to the requirement of high accuracy as it affects the detection of all data. The tradeoff between the decoding accuracy and the communication overhead is non-trivial but out of the scope of the paper and leaves for future work. 

\vspace{-10pt}\section{Simulation Results}\label{simulation}
The default simulation settings are as follows. The numbers of antennas are $N_{r} =4, N_{t} = 2$. The channel follows  block fading channel model and channel coefficients  i.i.d. $\cal{CN}$(0,1) r.v.. The noise follows the same distribution.  The constellation  size  and symbol length are $L = 8$ and  $T = 4$. 
\begin{figure}[t]
  \centering
\includegraphics[width=0.45\textwidth]{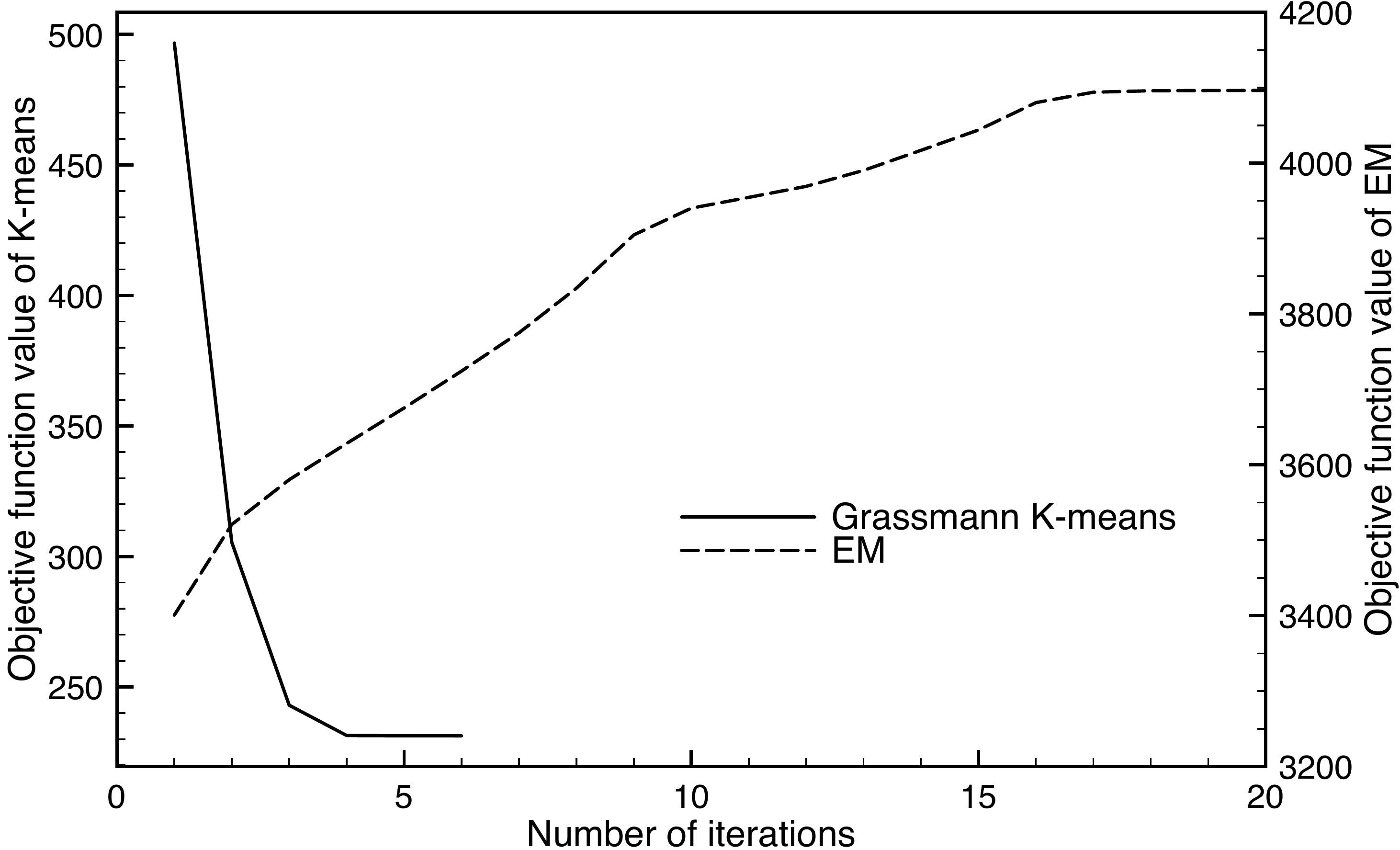} 
     \caption{Convergence-rate comparison between  K-means and EM algorithms for Grassmann constellation detection.}
  \vspace{-2mm}
  \label{Fig:Covergence}
\end{figure}

Consider the equivalence of  Grassmann K-means and EM algorithms derived in Section~\ref{bridge}. Their convergence rates  are compared in Fig.~\ref{Fig:Covergence}. One can observe that the former  converges faster than the latter. This aligns with the discussion in Section~\ref{convergence speed} and  confirms the advantage of the proposed data-clustering approach for Grassmann constellation detection.

\begin{figure*}[tt]
  \centering
  \subfigure[Effect of SNR]{\label{Fig:Effect of SNR}\includegraphics[width=0.45\textwidth]{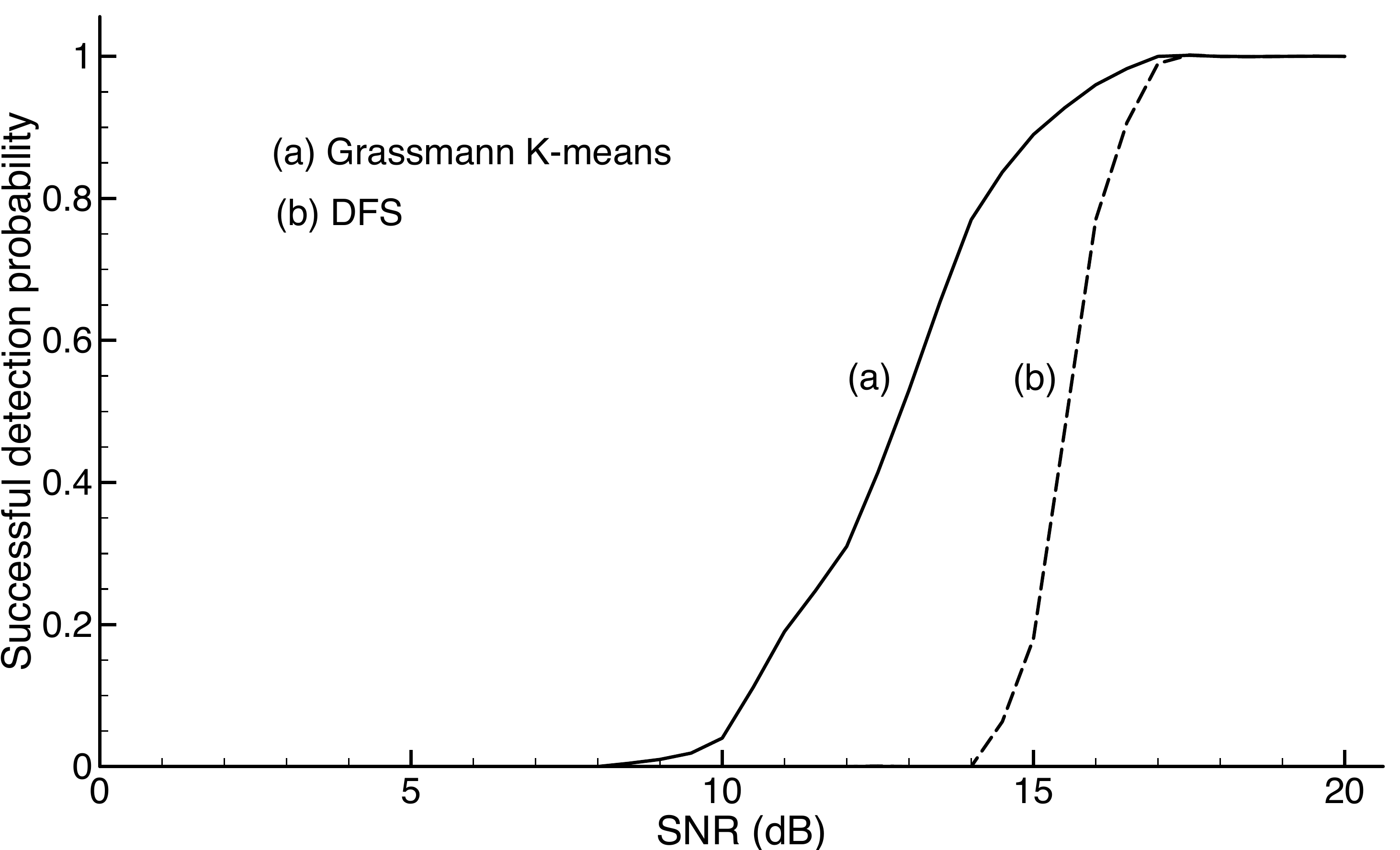}}
    \hspace{0.35in}
  \subfigure[Effect of constellation size]{\label{Fig:Effect of the constellation size}\includegraphics[width=0.45\textwidth]{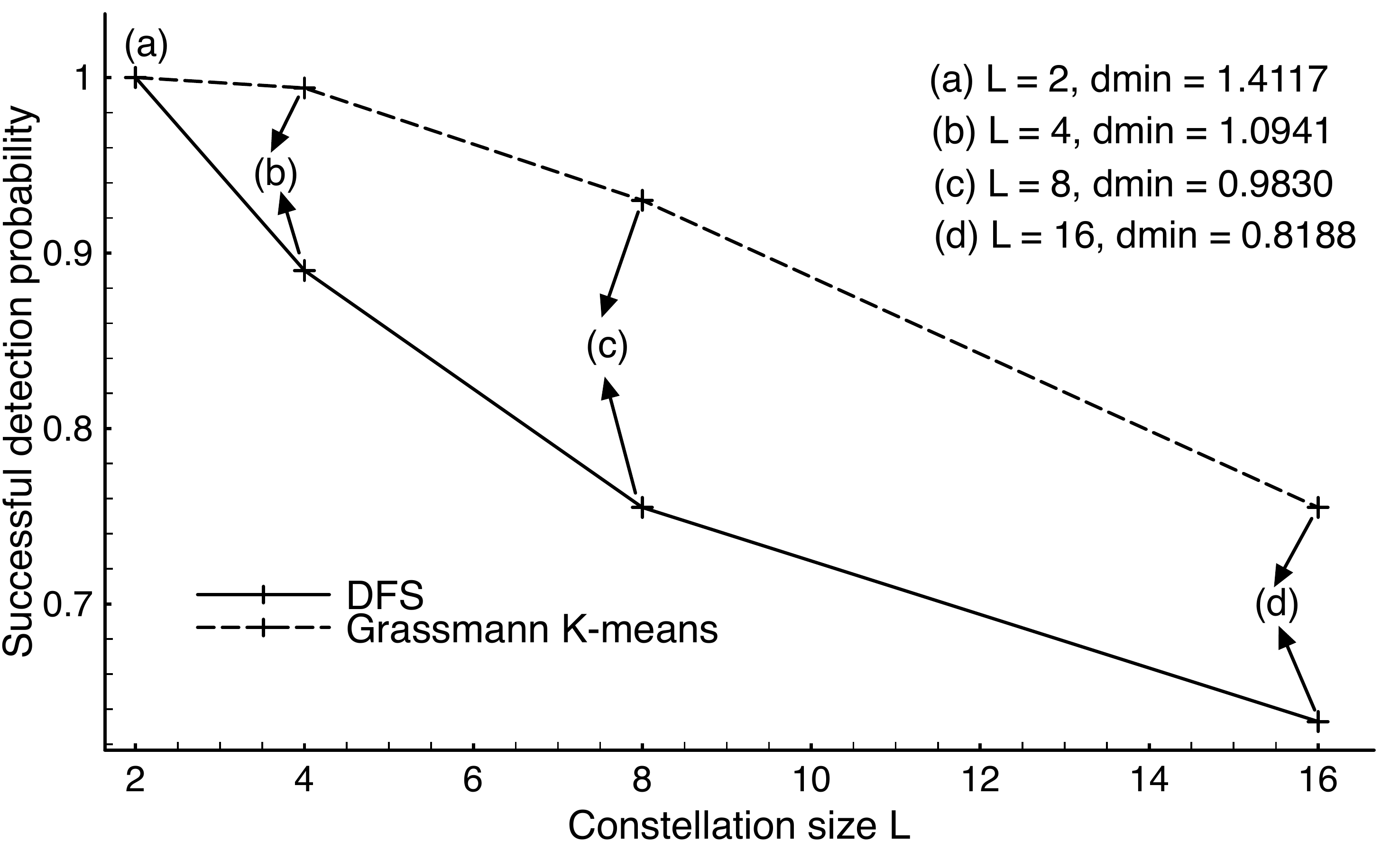}}
  \subfigure[Effect of dataset size]{\label{Fig:Effect of Data Size}\includegraphics[width=0.45\textwidth]{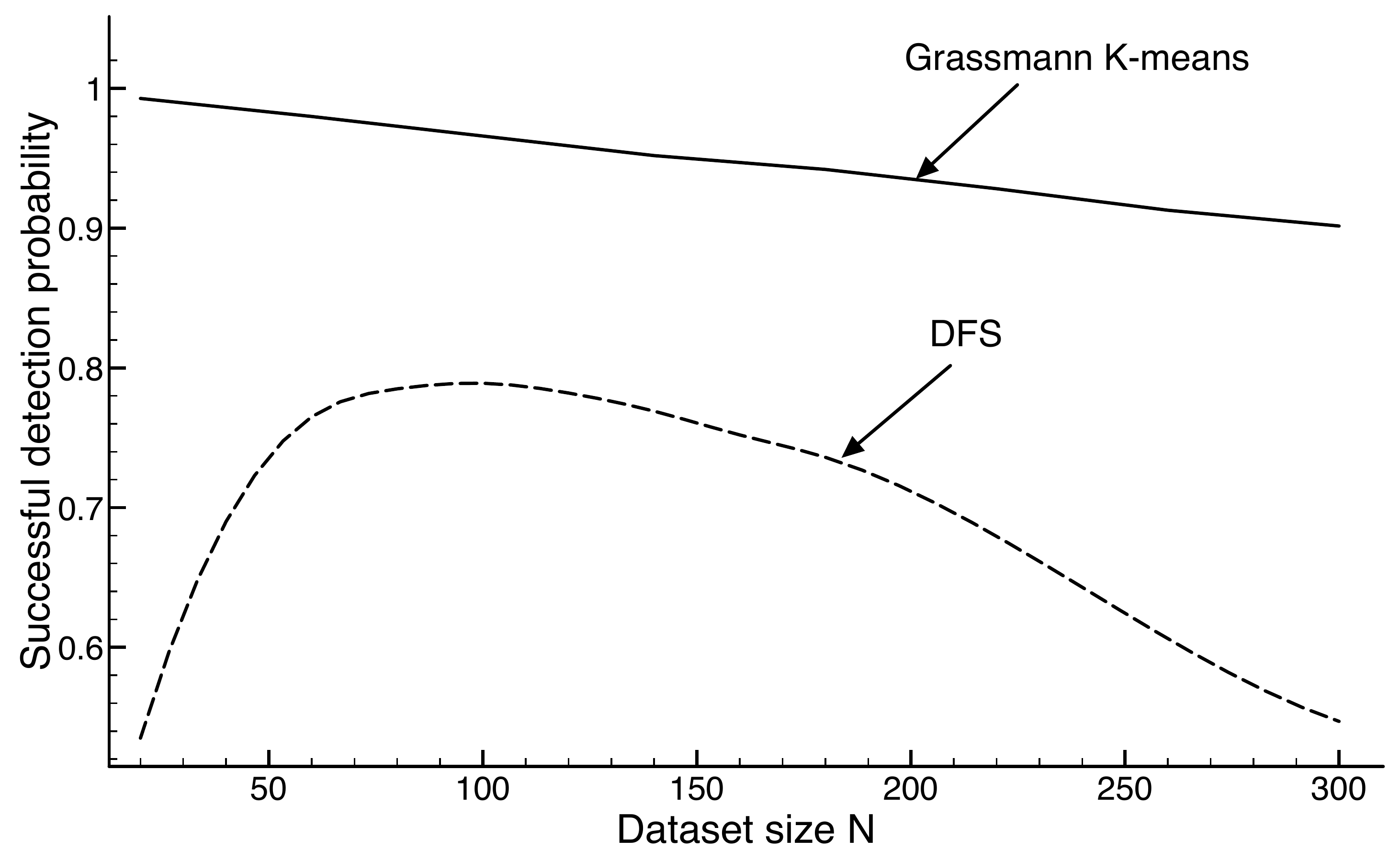}}
    \hspace{0.35in}
      \subfigure[Effect of DFS threshold]{\label{Fig:Effect of the threshold}\includegraphics[width=0.45\textwidth]{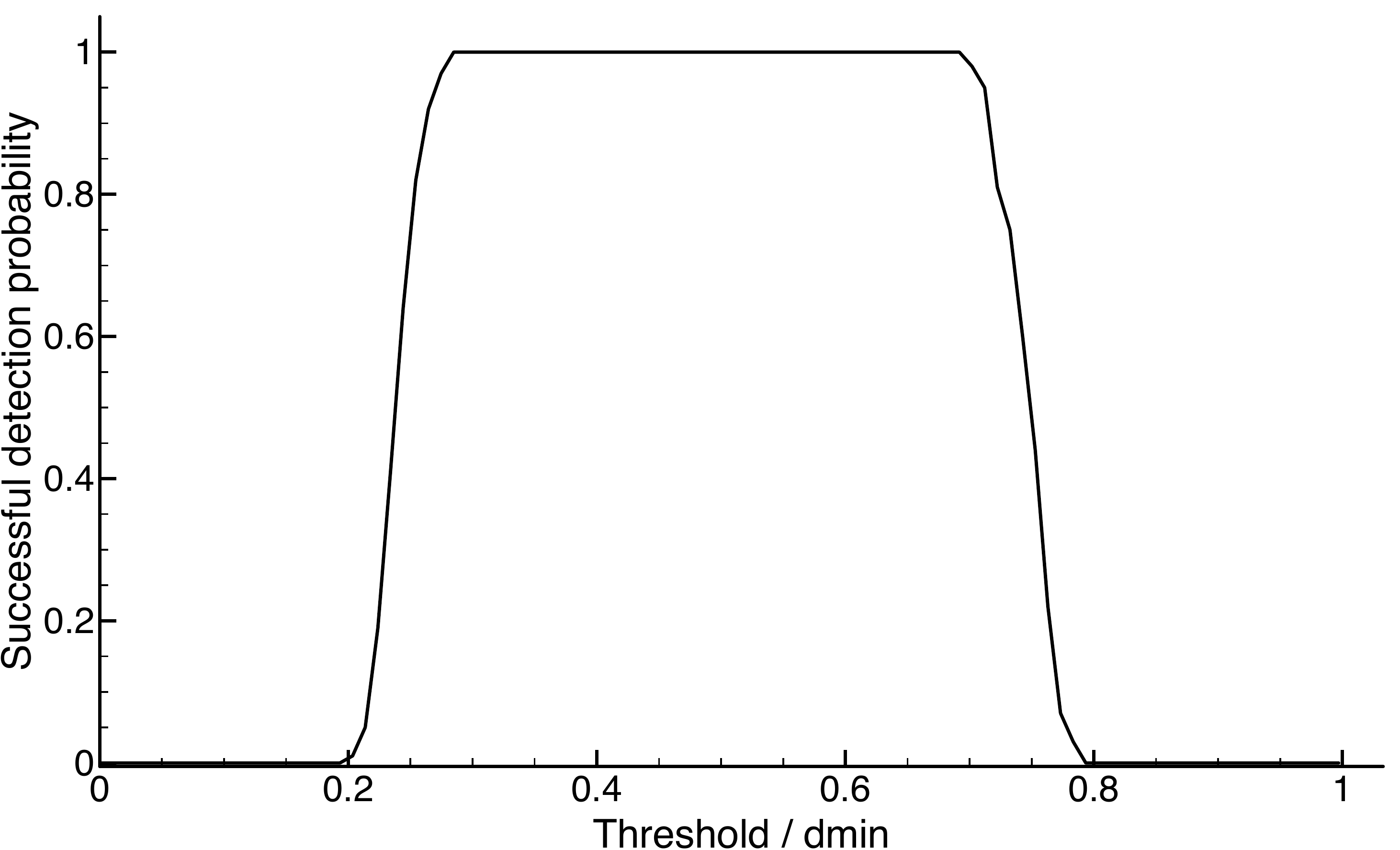}}
      \caption{Comparison of Grassmann constellation detection with and without knowledge of constellation size and the effects of parameters. }
        \label{fig_dfs_parameter}
  \vspace{-3mm}
\end{figure*}

In Fig.~\ref{fig_dfs_parameter}, we compare the performance of Grassmann constellation detection with and without the prior  knowledge of constellation size $L$, which are implemented using the K-means and DFS algorithms respectively.  Furthermore, each of key parameters is  varied to demonstrate its effect on the detection performance and thereby corroborate the analytical results. Define the \emph{successful detection probability} as the probability that the received symbols are correctly clustered according to their corresponding transmitted codewords. Using this metric for measuring the detection performance and 
by observing Fig.~\ref{fig_dfs_parameter}(a)-(c), the K-means is observed to substantially outperform the DFS, showing the value of the prior knowledge. 
Next, comparing Fig.~\ref{Fig:Effect of SNR} and \ref{Fig:Effect of the constellation size} reveals that the detection performance can be monotonically improved by increasing the  SNR or  reducing the constellation size  $L$, which agrees with the insights from the  analysis. On the other hand, as observed from Fig.~\ref{Fig:Effect of Data Size}, increasing the dataset size $N$ can have opposite effects on DFS  performance but continuously degrades the K-means performance. The reason is revealed in the  analysis: large $N$  improves the intra-cluster connectivity of DFS but degrades its inter-cluster separability while K-means performance only concerns separability.  In particular, the \emph{linear} decay rate  of success detection probability for K-means is predicted in \eqref{inter-cluster-exp-kmean}. Last, Fig.~\ref{Fig:Effect of the threshold} shows the sensitivity of the DFS performance towards the changes on  the DFS threshold and thus its optimization is important, which agrees with the analysis in Section \ref{DFS detection}.

\vspace{-10pt}\section{Concluding Remarks}\label{Section:Conclusion}
We  have proposed  an approach of automatic recognition of Grassmann constellations and developed an analytical framework for performance analysis. 
The work makes contributions to next-generation intelligent radios and opens up several interesting directions for further research including multiuser  constellation detection and detection using more complex machine learning tools such as deep learning. 

\noindent {\bf Acknowledgement:} Comments from Dr. Jun Zhang, Dr. Rahul Vaze and Dr. Sheng Yang have led to substantial improvements of this work. 

\vspace{-10pt}
\appendix

\vspace{-10pt}\subsection{Proof of Lemma \ref{single bound fast fading}}\label{proof:assignment criteria}
According to \eqref{decomposition_Y}, one can decompose the eigenspace of the received signal as:
\begin{align}\label{eigenspace_decomposition}
\bY^{(i)} = \bU^{(i)}_Y\boldsymbol{\Sigma}^{(i)}_Y(\bV^{(i)}_Y)^{H} + \bU^{(i)}_W\boldsymbol{\Sigma}^{(i)}_W(\bV^{(i)}_W)^{H}.		
\end{align}
where the first term captures the dominant signal subspace while the second one corresponds to the noise subspace. 
In the high SNR regime, the noise is  negligible and we have the following result
\begin{align}\label{argument_square}
\bY^{(i)} = \bU^{(i)}_Y\boldsymbol{\Sigma}^{(i)}_Y(\bV^{(i)}_Y)^{H}, \qquad \rho \rightarrow \infty.
\end{align}
It follows that 
\begin{equation}
\text{tr}\l\{ (\bY^{(i)})^{H}\hat{\boldsymbol\mu}_{j}\hat{\boldsymbol\mu}^{H}_{j}\bY^{(i)}  \r\} \longrightarrow  \text{tr}\l\{ (\boldsymbol\Sigma^{(i)}_{Y})^{2}(\bU^{(i)}_{Y})^{H}\hat{\boldsymbol\mu}_{j}\hat{\boldsymbol\mu}^{H}_{j}{\bU}^{(i)}_{Y}     \r\}, \qquad \rho \rightarrow \infty.\label{lema: hard assignment substitution}
\end{equation}
With $\bb_{k}$ denoting  the $k$th column of the matrix $\hat{\boldsymbol\mu}^{H}_{j}\bU^{(i)}_{Y}$ and  $\l\{\sigma^{(i)}_{k}\r\}_{k =1}^{N_t}$  singular values of $\boldsymbol\Sigma^{(i)}_{Y}$,
\begin{equation}
\text{tr}\l\{ (\boldsymbol\Sigma^{(i)}_{Y})^{2}(\bU^{(i)}_{Y})^{H}\hat{\boldsymbol\mu}_{j}\hat{\boldsymbol\mu}^{H}_{j}{\bU}^{(i)}_{Y}     \r\} = \sum_{k=1}^{N_t} \(\sigma^{(i)}_{k}\)^{2}{\| \bb_{k}\|}^{2}\label{quote_lowerbound_lem}.
\end{equation}
By replacing  $\l\{\sigma^{(i)}_{k}\r\}$ with the largest singular value  denoted as $\sigma^{(i)}_{1}$, 
\begin{align}
\text{tr}\l\{ (\bY^{(i)})^{H}\hat{\boldsymbol\mu}_{j}\hat{\boldsymbol\mu}^{H}_{j}\bY^{(i)}  \r\} 
 \leq \(\sigma^{(i)}_{1}\)^2 \text{tr}\l\{ (\bU^{(i)}_{Y})^{H}\hat{\boldsymbol\mu}_{j}\hat{\boldsymbol\mu}^{H}_{j}{\bU}^{(i)}_{Y}     \r\}, \qquad \rho \rightarrow \infty.\label{lema: hard assignment upper bound}
\end{align}
Similarly, the lower bound of $\text{tr}\l\{ (\bY^{(i)})^{H}\hat{\boldsymbol\mu}_{j}\hat{\boldsymbol\mu}^{H}_{j}\bY^{(i)}  \r\}$ can be obtained by replacing $\l\{\sigma^{(i)}_{k}\r\}$ in \eqref{quote_lowerbound_lem} with the smallest singular value denoted as $\sigma^{(i)}_{N_t}$:
\begin{align}
\text{tr}\l\{ (\bY^{(i)})^{H}\hat{\boldsymbol\mu}_{j}\hat{\boldsymbol\mu}^{H}_{j}\bY^{(i)}  \r\}  \geq  \(\sigma^{(i)}_{N_t}\)^2 \text{tr}\l\{ (\bU^{(i)}_{Y})^{H}\hat{\boldsymbol\mu}_{j}\hat{\boldsymbol\mu}^{H}_{j}{\bU}^{(i)}_{Y}     \r\}, \qquad \rho \rightarrow \infty.
\end{align}
Given that  $\bU^{(i)}_Y =\boldsymbol\Upsilon^{(i)}$, 
\begin{align}
\(\sigma^{(i)}_{N_t}\)^2 \text{tr}\l\{ (\boldsymbol\Upsilon^{(i)})^{H}\hat{\boldsymbol\mu}_{j}\hat{\boldsymbol\mu}^{H}_{j}\boldsymbol\Upsilon^{(i)}     \r\}     &\leq \text{tr}\l\{ (\bY^{(i)})^{H}\hat{\boldsymbol\mu}_{j}\hat{\boldsymbol\mu}^{H}_{j}\bY^{(i)}  \r\} \leq \(\sigma^{(i)}_{1}\)^2 \text{tr}\l\{ (\boldsymbol\Upsilon^{(i)})^{H}\hat{\boldsymbol\mu}_{j}\hat{\boldsymbol\mu}^{H}_{j}\boldsymbol\Upsilon^{(i)}     \r\}   \label{bounds_appendix}.
\end{align}
Rewriting the bounds in \eqref{bounds_appendix}  in terms of  Procrustes distance defined in~\eqref{eq: procrustes_dist} gives  the desired result. 

\vspace{-10pt}\subsection{Proof of Lemma \ref{Lem: SufficientDataSize}}\label{proof: SufficientDataSize}
Let  $p_{\ell}$ denote  the joint probability of  two events, namely  ${\cal A}$: a symbol generated from the $\ell$-th codeword and $\cal B$: a symbol is assigned to cluster $\ell$.  One can easily see that $N_{\ell} \geq p_{\ell}N$. Therefore, as long as $p_{\ell}$ is bounded by some strictly positive value, the statement holds. To show this, according to the equal-probability codeword assumption, we have $p({\cal A}) = \frac{1}{L}$, and by definition we also have $p({\cal B | A}) \geq  p(d_p(\boldsymbol\Upsilon,\boldsymbol\mu) \leq \frac{d_{\min}}{2})$. It follows that $p_{\ell} \geq \frac{1}{L}p(d_p(\boldsymbol\Upsilon,\boldsymbol\mu) \leq \frac{d_{\min}}{2})$, where $p(d_p(\boldsymbol\Upsilon,\boldsymbol\mu) \leq \frac{d_{\min}}{2})$ can be directly derived from Lemma \ref{lemma: coverage of a typical cluster}. Thus $p_{\ell}$ is indeed strictly positive. Consequently, $N \rightarrow \infty$ can lead to $N_{\ell} \rightarrow \infty$, completing the proof.

\vspace{-10pt}\subsection{Proof of Lemma \ref{fast_fading_infinite}}\label{proof:fast_fading_infinite}

By substituting  $\boldsymbol\mu^{*}_{\ell}\bH^{(i)}+\sqrt{\frac{N_t}{\rho T}}\bW^{(i)}$ into $\bY^{(i)}$, $\frac{1}{N_{\ell}}\underset{{i \in {\mathcal C}_{\ell}}} \sum\text{tr}\l\{ (\bY^{(i)})^{H}\boldsymbol\mu_{\ell}\boldsymbol\mu^{H}_{\ell}\bY^{(i)}  \r\}$ can be rewritten as  
%$\frac{1}{N_c}\underset{{i \in {\mathcal S}_{\ell}}} \sum\text{tr}\l\{ \bY^{(i)}\boldsymbol\mu^{H}_{\ell}\boldsymbol\mu_{\ell}(\bY^{(i)})^{H}  \r\}$ is equal to
\begin{equation}
\frac{1}{N_{\ell}}\underset{{i \in {\mathcal C}_{\ell}}} \sum\text{tr}\l\{ \(\boldsymbol\mu^{*}_{\ell}\bH^{(i)}+\sqrt{\frac{N_t}{\rho T}}\bW^{(i)}\){\(\boldsymbol\mu^{*}_{\ell}\bH^{(i)}+\sqrt{\frac{N_t}{\rho T}}\bW^{(i)}\)^{H}}\boldsymbol\mu_{\ell}\boldsymbol\mu^{H}_{\ell}  \r\}\label{eq: cited_in_slow_fading}.
\end{equation}
Using the law of large numbers, as  $N_{\ell} \rightarrow \infty$, $\frac{1}{N_{\ell}}\underset{{i \in {\mathcal C}_{\ell}}} \sum\text{tr}\l\{ (\bY^{(i)})^{H}\boldsymbol\mu_{\ell}\boldsymbol\mu^{H}_{\ell}\bY^{(i)}  \r\}$ can thus be simplified as 
\begin{equation}
\frac{1}{N_{\ell}}\underset{{i \in {\mathcal C}_{\ell}}} \sum\text{tr}\l\{ (\bY^{(i)})^{H}\boldsymbol\mu_{\ell}\boldsymbol\mu^{H}_{\ell}\bY^{(i)}  \r\} \longrightarrow \text{tr}\l\{  \boldsymbol\mu^{*}_{\ell}(\boldsymbol\mu^{*}_{\ell})^{H}\boldsymbol\mu_{\ell}\boldsymbol\mu^{H}_{\ell} \r \} + \frac{N^{2}_{t}}{\rho T}. \label{quote begin}
\end{equation}
Let  $\bQ^{(i)}$ denote  the unitary matrix, 
\begin{equation}
\text{tr}\l\{  \boldsymbol\mu^{*}_{\ell}(\boldsymbol\mu^{*}_{\ell})^{H}\boldsymbol\mu_{\ell}\boldsymbol\mu^{H}_{\ell}  \r\}=\text{tr}\l\{ \frac{1}{N_{\ell}}\(\underset{{i \in {\mathcal C}_{\ell}}} \sum{(\boldsymbol\mu^{*}_{\ell}\bQ^{(i)})(\boldsymbol\mu^{*}_{\ell}\bQ^{(i)})^{H}}\)\boldsymbol\mu_{\ell}\boldsymbol\mu^{H}_{\ell}  \r\}. 
\end{equation}
Moreover, as $\rho \rightarrow \infty$, the noise effect is negligible, resulting in  $  \bU^{(i)}_Y \rightarrow \boldsymbol\mu^{*}_{\ell}\bQ^{(i)} = \boldsymbol\Upsilon^{(i)}$. This can be interpreted as  as an approximation of the column space spanned by the received signal $\bY^{(i)}$. Thereby, we have the following result.
\begin{equation}
\frac{1}{N_{\ell}}\underset{{i \in {\mathcal C}_{\ell}}} \sum\text{tr}\l\{ (\bY^{(i)})^{H}\boldsymbol\mu_{\ell}\boldsymbol\mu^{H}_{\ell}\bY^{(i)}  \r\} \longrightarrow \frac{1}{N_{\ell}}\underset{{i \in {\mathcal C}_{\ell}}} \sum\text{tr}\l\{ \boldsymbol\Upsilon^{(i)}(\boldsymbol\Upsilon^{(i)})^{H}\boldsymbol\mu_{\ell}\boldsymbol\mu^{H}_{\ell}     \r\}, \qquad \rho \rightarrow \infty. \label{quote end}
\end{equation}
This  completes the proof. 

\vspace{-4mm}
\subsection{Proof of monotonous decreasing property of $p(r)$}\label{proof: monoton_property}
Note that the first term in \eqref{eq_location}, i.e. $\frac{\eta^{-1}_{_D}(\frac{\gamma_0}{4})^{D}}{r^{D}-(r-\frac{\gamma_0}{2})^{D}}$, decreases monotonically with respect to $r$, hence, it is sufficient to prove the monotonically decreasing characteristics of the second term for $r \geq \frac{\gamma_0}{2}$. By defining $f(r) = \Gamma\(\frac{D}{2},\frac{\rho T{\bar\lambda^2}(r-\frac{\gamma_0}{2})^2}{2N_t}\)-\Gamma\(\frac{D}{2},\frac{\rho T{\bar\lambda}^2r^2}{2N_t}\)$ and setting its first derivative to $0$, the following equality holds
\begin{align}
\frac{\gamma_0}{2r} = 1-e^{-\frac{\rho T {\bar\lambda}^2\gamma_0(r-\frac{\gamma_0}{4})}{2N_t(D-1)}}.
\end{align}
Observe that as $\rho \to \infty$, $e^{-\frac{\rho T {\bar\lambda}^2\gamma_0(r-\frac{\gamma_0}{4})}{2N_t(D-1)}} \to 0$, we thus have $r = \frac{\gamma_0}{2}$, which implies that $f(r)$ decreases monotonically for $r \geq \frac{\gamma_0}{2}$.
We complete the whole proof.

\subsection{Computation of $p_{\min}$}\label{proof: p_min_calculation}
By substituting $x$ in \eqref{eq_location} with $\frac{a}{\sqrt{\rho}}$, we have
\begin{align}\label{p_min_exact}
p_{\min}  \! =\! \frac{\eta^{-1}_{_D}(\frac{\gamma_0}{4})^{D}}{{\Gamma(\frac{D}{2})}\((\frac{a}{\sqrt{\rho}})^{D}\!-\!(\frac{a}{\sqrt{\rho}}-\frac{\gamma_0}{2})^{D}\)} \l\{ \Gamma\(\frac{D}{2},\frac{\rho T{\bar\lambda^2}\(\frac{a}{\sqrt{\rho}}-\frac{\gamma_0}{2}\)^2}{2N_t}\)-\Gamma\(\frac{D}{2},\frac{ T{\bar\lambda}^2a^2}{2N_t}\)\r\}.
\end{align}
Next, consider $\Gamma\(\frac{D}{2},\frac{\rho T{\bar\lambda^2}\(\frac{a}{\sqrt{\rho}}-\frac{\gamma_0}{2}\)^2}{2N_t}\)=\Gamma\(\frac{D}{2},\frac{a^2T{\bar\lambda^2} + \rho T{\bar\lambda^2} \frac{{\gamma_0}^2}{4} - a\gamma_0T{\bar\lambda^2}\sqrt{\rho}}{2N_t}\)$. Under the assumption that $\frac{\gamma_0}{2} \ll \frac{a}{\sqrt{\rho}}$, we have $\frac{\rho T{\bar\lambda^2} \frac{{\gamma_0}^2}{4}}{a\gamma_0T{\bar\lambda^2}\sqrt{\rho}} = \frac{1}{2}\frac{\frac{\gamma_0}{2}}{\frac{a}{\sqrt{\rho}}} \to 0$, holds. Ignoring the high-order term $\rho T{\bar\lambda^2} \frac{{\gamma_0}^2}{4}$, one can have
\begin{align}\label{approx_0}
\Gamma\(\frac{D}{2},\frac{\rho T{\bar\lambda^2}\(\frac{a}{\sqrt{\rho}}-\frac{\gamma_0}{2}\)^2}{2N_t}\) &  = \Gamma\(\frac{D}{2},\frac{a^2T{\bar\lambda^2} - a\gamma_0T{\bar\lambda^2}\sqrt{\rho}}{2N_t}\) + o({\gamma_0}^{D}{\rho}^{\frac{D}{2}})\nn\\
& =  \Gamma\(\frac{D}{2},\frac{a^2T{\bar\lambda^2} }{2N_t}\) + \int_{\frac{a^2T{\bar\lambda^2} - a\gamma_0T{\bar\lambda^2}\sqrt{\rho}}{2N_t}}^{\frac{a^2T{\bar\lambda^2}}{2N_t}} x^{\frac{D}{2}-1}e^{-x} dx + o({\gamma_0}^{D}{\rho}^{\frac{D}{2}}).
\end{align}
Realizing the fact that $\frac{a\gamma_0T{\bar\lambda^2}\sqrt{\rho}}{a^2T{\bar\lambda^2}} = \frac{\gamma_0}{\frac{a}{\sqrt{\rho}}} \to 0$, the second term of \eqref{approx_0} can be rewritten as
\begin{align}\label{approx_1}
\int_{\frac{a^2T{\bar\lambda^2} - a\gamma_0T{\bar\lambda^2}\sqrt{\rho}}{2N_t}}^{\frac{a^2T{\bar\lambda^2}}{2N_t}} x^{\frac{D}{2}-1}e^{-x} dx = x_0^{\frac{D}{2}-1}e^{-x_0}\Delta{x_0} + o(\gamma_0\sqrt{\rho}),
\end{align}
where $\Delta{x_0} = \frac{a\gamma_0T{\bar\lambda^2}\sqrt{\rho}}{2N_t}$, $x_0 = \frac{a^2T{\bar\lambda^2}}{2N_t}$. 
%Thereby, we have
%\begin{align}\label{approx_gamma}
%\Gamma\(\frac{D}{2},\frac{\rho T{\bar\lambda^2}\(\frac{a}{\sqrt{\rho}}-\frac{\gamma_0}{2}\)^2}{2N_t}\)-\Gamma\(\frac{D}{2},\frac{ T{\bar\lambda}^2a^2}{2N_t}\) \approx \int_{a^2T{\bar\lambda^2} - a\gamma_0T{\bar\lambda^2}\sqrt{\rho}}^{a^2T{\bar\lambda^2}} x^{\frac{D}{2}-1}e^{-x} dx.
%\end{align}
Moreover, since
${x}^D-\({x}-\frac{\gamma_0}{2}\)^D \!=\! \frac{D\gamma_0}{2}x^{D-1} + o(\gamma_0x^{D-1})$ for $\frac{\gamma_0}{2} \ll x$, which can be directly proved using \emph{Taylor expansion}, we thus have 
\begin{align}\label{approx_2}
\(\frac{a}{\sqrt{\rho}}\)^D-\({\frac{a}{\sqrt{\rho}}}-\frac{\gamma_0}{2}\)^D = \frac{D\gamma_0}{2}\(\frac{a}{\sqrt{\rho}}\)^{D-1} + o(\gamma_0{\rho}^{-\frac{D-1}{2}}).
\end{align}
By integrating the above approximations, i.e. \eqref{approx_0} $\sim$ \eqref{approx_2}, into \eqref{p_min_exact}, the whole proof is completed.
\vspace{-4mm}

\bibliography{reference.bib}
\bibliographystyle{IEEEtran}

\end{document}